\documentclass{article}
\usepackage[letterpaper,top=2cm,bottom=2cm,left=2cm,right=2cm,marginparwidth=1.75cm]{geometry}

\usepackage{mathtools}
\usepackage{subfig}
\usepackage{graphicx}
\usepackage{amssymb}
\usepackage{hyperref}

\title{Survey of Methods for Solving Systems of Nonlinear Equations, Part II: Optimization Based Approaches}
\author{Ilias S. Kotsireas \\
ikotsire@wlu.ca \\
Wilfrid Laurier University  \\
Canada
\and
Panos M. Pardalos  \\
ppardalos@toxeus.org  \\
Toxeus Systems LLC \\
Orlando, Florida, USA
\and
Alexander Semenov \\
asemenov@ufl.edu \\
University of Florida \\
Gainesville, Florida, USA
\and
William T. Trevena \\
wtrevena@ufl.edu \\
University of Florida \\
Gainesville, Florida, USA
\and
Michael N. Vrahatis \\
vrahatis@math.upatras.gr \\
University of Patras \\
Patras, Greece
}
\begin{document}

%\renewcommand{\shortauthors}{I. S. Kotsireas {\em et al.}}
\maketitle
\begin{abstract}
    This paper presents a comprehensive survey of methods which can be utilized to search for solutions to systems of nonlinear equations (SNEs). Our objectives with this survey are to synthesize pertinent literature in this field by presenting a thorough description and analysis of the known methods capable of finding one or many solutions to SNEs, and to assist interested readers seeking to identify solution techniques which are well suited for solving the various classes of SNEs which one may encounter in real world applications.
  
  To accomplish these objectives, we present a multi-part survey. In part one, we focused on root-finding approaches which can be used to search for solutions to a SNE without transforming it into an optimization problem. In part two, we introduce the various transformations which have been utilized to transform a SNE into an optimization problem, and we discuss optimization algorithms which can then be used to search for solutions. We emphasize the important characteristics of each method, and we discuss promising directions for future research. In part three, we will present a robust quantitative comparative analysis of methods capable of searching for solutions to SNEs.
  
%   techniques which can be utilized to determine the number of solutions to a SNE within a bounded domain, and we then describe both optimization-based and non-optimization-based techniques which have been utilized in literature to search for solutions to SNEs. We emphasize the important characteristics of each method, and we conclude our survey with a discussion of promising directions for future research.
  
%   describe the limitations advantages and disadvantages of each method, as well as the classes of SNEs that each method is best suited for solving. We conclude our suvey with a discussion of promising directions for future research. %solution technique is best suited for. We highlight the advantages and disadvantages of each method, and we the surveyed methods, and we  the root finding methods

 % and highlights emerging applications of SNEs in real-world problems. 
%   We perform a comprehensive survey of the families of SNE solution methods including evolution strategies, genetic algorithms, particle swarm optimization (PSO), homotopy methods, the continuous variable neighborhood search (C-VNS) heuristic, and the the continuous greedy randomized adaptive search procedure (C-GRASP) heuristic.
  %We conclude our paper with a discussion of promising future research directions. Our objective is to present a thorough analysis of the state of the art methods capable of finding one or multiple solutions to SNEs, and to synthesize literature in the field to highlight promising directions for future research efforts.
\end{abstract}

% %  <concept>
% %   <concept_id>10010520.10010553.10010562</concept_id>
% %   <concept_desc>Computer systems organization~Embedded systems</concept_desc>
% %   <concept_significance>500</concept_significance>
% %  </concept>
% %  <concept>
% %   <concept_id>10010520.10010575.10010755</concept_id>
% %   <concept_desc>Computer systems organization~Redundancy</concept_desc>
% %   <concept_significance>300</concept_significance>
% %  </concept>
% %  <concept>
% %   <concept_id>10010520.10010553.10010554</concept_id>
% %   <concept_desc>Computer systems organization~Robotics</concept_desc>
% %   <concept_significance>100</concept_significance>
% %  </concept>
% %  <concept>
% %   <concept_id>10003033.10003083.10003095</concept_id>
% %   <concept_desc>Networks~Network reliability</concept_desc>
% %   <concept_significance>100</concept_significance>
% %  </concept>
% \ccsdesc[500]{}



\maketitle

% \tableofcontents

\section{Introduction}
This is the second part of a survey on methods for finding one or many real solutions to a {\em system of nonlinear equations}\/ (SNE):
% \begin{equation} \label{eq1}
% \begin{split}
% A & = \frac{\pi r^2}{2} \\
%  & = \frac{1}{2} \pi r^2
% \end{split}
% \end{equation}
\begin{equation}\label{eq:1}
{F}_m(x) = {\mathit \Theta}_{m} \equiv (0, 0, \ldots, 0)^{\top} \quad \Longleftrightarrow \quad \left\{\begin{aligned}
    &f_1(x_1,x_2,\ldots,x_n)=0, \\[0.1cm]
    &f_2(x_1,x_2,\ldots,x_n)=0, \\
    &\kern1.45cm \vdots \\
%    \vdotswithin{=} \\
    &f_m(x_1,x_2,\ldots,x_n)=0,
\end{aligned} \right.
\end{equation}
% Below I changed a_n,b_n to a_m,b_m so we have n equations and m variables
where ${F}_m=(f_1,f_2,\ldots, f_m): {\mathcal D}_n \subset {\mathbb R}^n
\to {\mathbb R}^m$,
%$x \in S = [a_1,b_1] \times [a_2,b_2]\times \cdots \times [a_n,b_n] \subset \mathbb{R}^n$ 
where $f_1,f_2,\ldots,f_m$ are real-valued continuous or continuously differentiable functions on the domain ${\mathcal D}_n$, and where at least one of $f_1,f_2,\ldots,f_m$ is nonlinear. For example, consider the system of transcendental equations
\begin{equation}\label{eq:00}
    {F}_2(x) = {\mathit \Theta}_{2} \equiv (0, 0)^{\top} \quad \Longleftrightarrow \quad \left\{\begin{aligned}
    f_1(x_1,x_2)&=x_1-x_1\sin(x_1+5x_2)-x_2\cos(5x_1-x_2)=0, \\[0.1cm]
    f_2(x_1,x_2)&=x_2-x_2\sin(5x_1-3x_2)+x_1\cos (3x_1+5x_2)=0,
\end{aligned}  \right.
\end{equation}
which is comprised of two transcendental equations of two unknowns (See Figure 1). 

\begin{figure}[h]\label{fig1}
  \centering
  \includegraphics[width=\linewidth]{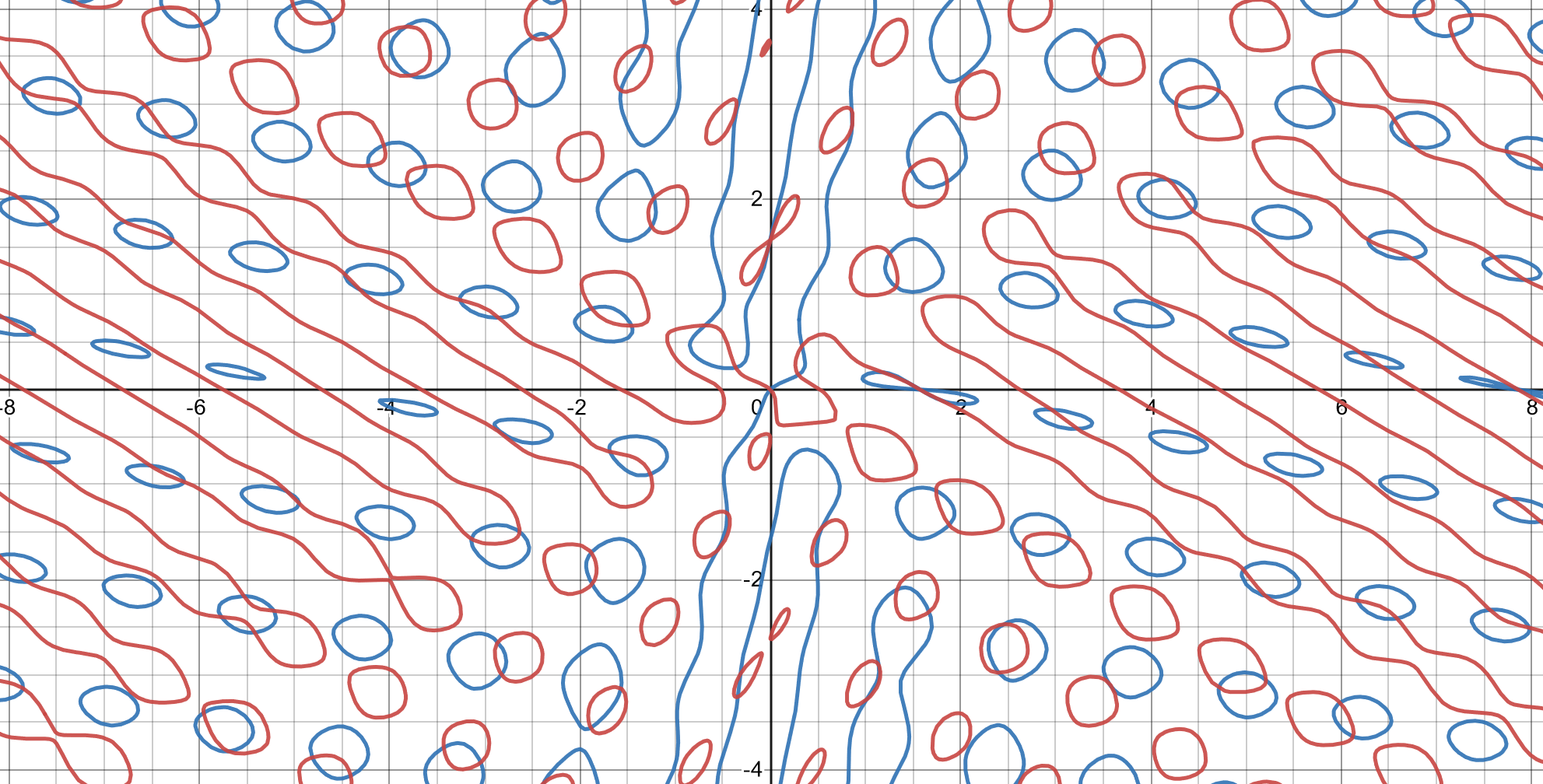}
  \caption{An example of a SNE with two transcendental equations of two unknowns as introduced by Eq.~(\ref{eq:00}): \\(Blue): $f_1(x_1,x_2) = x_1-x_1\sin(x_1+5x_2)-x_2\cos(5x_1-x_2)=0$; \ \ (Red): $f_2(x_1,x_2) = x_2-x_2\sin(5x_1-3x_2)+x_1\cos (3x_1+5x_2)=0$. Solutions to this SNE are defined as the points where the blue and red contours intersect. Finding all of the points within a certain region which satisfy both equations is a challenging task. %Graph generated by Desmos Graphing Calculator. (\url{https://www.desmos.com/calculator}).
  }
%   \Description{A woman and a girl in white dresses sit in an open car.}
\end{figure}

Finding one or more solutions to a SNE is a challenging and ubiquitous task faced in many fields including chemistry \cite{FLOUDAS2000125,SACCO20115424,Holstad1999}, chemical engineering \cite{Jimenez-Islas2013},
automotive steering \cite{HENDERSON2010551},
power flow \cite{Chiang2014, mehta2016numerical}, large-scale integrated circuit designs \cite{Chiang2018},
% PAPER \cite{osti_6449249} WAS REFERENCED IN \cite{Martinez1994} BUT COULD NOT BE FOUND
% aeronautics \cite{osti_6449249}, 
% PAPER \cite{Martinez1994 MENTIONS ELECTRICAL ENGINEERING / POWER FLOW PROBLEMS BUT DOESNT GIVE AN EXAMPLE
% electrical engineering \cite{Martinez1994},
climate modeling \cite{yang2010fully},
% PAPER \cite{MO20091877} ONLY MENTIONS PETROLEUM GEOLOGICAL PROSPECTING, BUT DOESNT GIVE AN EXPLICIT PROBLEM
% petroleum geological prospecting \cite{MO20091877}, 
materials engineering \cite{Schneider2019}, robotics \cite{ZHANG2006,Lafmejani20151,cox2015robotics,Ji2003KinematicsAO}, 
% PAPER \cite{meschke2011computational} MAY NOT DISCUSS SNEs 
%computational mechanics \cite{meschke2011computational}, 
% PAPER kuznetsov1997mathematical WAS ORIGINALLY IN NUCLEAR ENGINEERING BUT SEEMS TO DISCUSS SYSTEMS OF NONLINEAR DIFFERENTIAL EQUATIONS
nuclear engineering \cite{cht2016modification}, 
% PAPER \cite{6968423} ONLY MENTIONS MEDICINE, BUT DOESNT GIVE AN EXPLICIT PROBLEM
% medicine \cite{6968423},
image restoration \cite{aji2020modified}, 
% COULDNT FIND \cite{RePEccmtpumathpuma1993v004pp0199-0209}, EVEN ON LIBGEN
% economics \cite{RePEccmtpumathpuma1993v004pp0199-0209}, 
% PAPER \cite{petkovic2016some} ONLY MENTIONS ECONOMICS BUT DOESNT GIVE AN EXPLICIT EXAMPLE
% economics \cite{petkovic2016some}, 
protein interaction networks \cite{Chiang2018}, neurophysiology \cite{verschelde1994}, economics \cite{Grosan2008}, 
finance \cite{golbabai2012}, applied mathematics \cite{zhang2009existence}, physics \cite{PhysRevLett.81.1195}, finding string vacua \cite{Mehta2011}, machine learning \cite{song2020nonlinear,pmlr-v108-cai20b}, and geodesy \cite{palancz2008dixon,Palancz2009} among others. 
%Furthermore, applications of SNEs expand into nonlinear programming as the first-order optimality conditions are characterized by a nonlinear system. 
The problem of solving even a system of polynomial equations has been proven to be NP-hard \cite{Jansson1998}.  Furthermore, it has also been proven \cite{Matiyasevich} that no general algorithm exists for determining whether an integer solution exists for a polynomial equation with a finite number of unknowns and only integer coefficients. The latter has been known as Hilbert's 10th problem.

\subsection{Notation / Scientific Style}
Throughout this paper, we utilize $x=(x_1, x_2, \ldots, x_n)^{\top} \in \mathcal{D}\subset \mathbb{R}^n$ to denote a real vector within the bounded domain $\mathcal{D}$. Furthermore, we utilize $x^*=(x_1^*, x_2^*, \ldots, x_n^*)^{\top} \in \mathcal{D}\subset \mathbb{R}^n$ to denote a real solution to a SNE such that all equations in the SNE are satisfied ($F_m(x^*)=0$). In an iterative method, we utilize $x^k=(x_1^k, x_2^k, \ldots,x_i^k,\ldots x_n^k)^{\top} \in \mathcal{D}\subset \mathbb{R}^n$ for $k=0,1,\ldots$ to denote the vector found during the $k-$th iteration of the iterative method. Here, $x^k_i$ denotes the $i-$th coordinate of the vector $x^k$. When we discuss optimization methods, 
%because $f_iinstead of the classical notation $f:\mathbb{R}^n\rightarrow\mathbb{R}$, 
we denote an objective function as $\varphi:\mathbb{R}^n\rightarrow\mathbb{R}$, and we denote the corresponding gradient as $\nabla \varphi(x)$. %$\varphi(x)$
%We utilize $a,\rho\in \mathbb{R}$ to denote constants.

\subsection{Terminology}
% \paragraph{} 
% \vspace{0.2cm}
% \noindent
% \textbf{Terminology.}
Although we refer to Eq.~ (\ref{eq:00}) as a \textit{system of nonlinear equations} (SNE), such systems have been referred to in a variety of different ways in literature. For example, articles \cite{khirallah2013solving,khirallah2012novel,koupaei2015,el-shorbagy2020,naidu2016solving,kuri2003} utilize the abbreviation ``SNLE" to refer to a \textit{system of nonlinear equations}, and article \cite{xiao2018solving} uses the abbreviation ``SoNE". 
Other papers refer to Eq.~ (\ref{eq:00}) as a \textit{nonlinear system of equations}, and use the abbreviations ``NSE" \cite{Geng2009,pei2019} and ``NLS"  \cite{Mousa2008}. Eq.~ (\ref{eq:00}) has also been referred to as a \textit{nonlinear equation system} (NES) \cite{Gong2017,6849952,Qin2015,Guo2020b,LIAO2020113261,Liao2020,Gao2020,Gong2020c,Gao2021,HE2019104796,WU2021106733,Gao2021b,Song2020,LIAO2020105312,Cheng2018}. The survey in article \cite{gong2021nonlinear} uses the terminology \textit{Nonlinear Equations} (NEs) to refer to a system of one or more nonlinear equations.

When $m>n$, a SNE can be referred to as an {\em overdetermined} SNE, and when $n>m$, a SNE can be referred to as an {\em underdetermined} SNE. When $m=n$, a SNE can be referred to as a \textit{square} SNE \cite{Ahookhosh2013}. Furthermore, a SNE is considered to be \textit{consistent} if a solution exists which satisfies all equations \cite{Gatilov2011PropertiesON}.

For a \textit{square} SNE, a {\em solution $x^* =(x_1^*, x_2^*, \ldots, x_n^*)^{\top}$ of the} SNE {\em ${F}_n(x) = {\mathit \Theta}_{n}$} or equivalently a {\em zero $x^*$ of the function ${F}_n(x)$}\/ or a {\em root $x^*$ of the function ${F}_n(x)$}\/ is called {\em simple}\/ if for the determinant of the corresponding {\em Jacobian matrix}:
\begin{equation}\label{jacobian}
 J_{F_{n}}(x) \equiv F'_n(x)_{ij} \equiv
\left\{\frac{\displaystyle\partial f_i(x)}{\displaystyle\partial x_j } \right\}_{ij} \equiv
\renewcommand{\tabcolsep}{2.5mm}
\renewcommand{\arraystretch}{1.6}
\left[ \begin{array}{cccc}
 \vspace*{0.1cm}
{\frac{\displaystyle\partial f_1(x)}{\displaystyle\partial x_1}} &
{\frac{\displaystyle\partial f_1(x)}{\displaystyle\partial x_2}} &
\cdots &
{\frac{\displaystyle\partial f_1(x)}{\displaystyle\partial x_{n}}}\\[0.2cm]
{\frac{\displaystyle\partial f_2(x)}{\displaystyle\partial x_1}} &
{\frac{\displaystyle\partial f_2(x)}{\displaystyle\partial x_2}} &
\cdots &
{\frac{\displaystyle\partial f_2(x)}{\displaystyle\partial x_{n}}}\\
 \vdots & \vdots & \ddots & \vdots\\[0.1cm]
{\frac{\displaystyle\partial f_n(x)}{\displaystyle\partial x_1}} &
{\frac{\displaystyle\partial f_n(x)}{\displaystyle\partial x_2}} &
\cdots &
{\frac{\displaystyle\partial f_n(x)}{\displaystyle\partial x_{n}}}  %\vspace*{0.1cm}
\end{array} \right],\vspace*{0.1cm}
\end{equation}%
at $x^*$ it holds that ${\rm det}\,J_{F_{n}}(x^*) \neq 0$, otherwise it is called {\em multiple}.
The problem of conservation
and decomposition of a multiple root into simple roots
in the case of
systems of homogeneous algebraic equations has been tackled in~\cite{TanabeV2006}.
This approach can be applied to high dimensional
CAD where it is sometimes required to compute
% the computation of
the  
intersection of several hypersurfaces
that are a perturbation of a set of original unperturbed hypersurfaces.
%{TanabeV2006}
%Tanab{\'e}, Susumu and Vrahatis, Michael N.
%On perturbation of roots of homogeneous algebraic systems,
%Mathematics of Computation, 75(255), pp.1383-1402, 2006.
 
%Also, %in this case, 
When $F_n$ satisfies the {\em monotonicity condition}:
\begin{equation}\label{monotonicity}
\bigl(F_n(x)-F_n(y)\bigr)^\top (x-y) \geqslant 0,\kern0.3cm \forall\, x,y\in \mathbb{R}^n,
\end{equation}
the corresponding SNE can be referred to as a \textit{system of monotone nonlinear equations}~\cite{Ullah2021}. Furthermore, $F_n$ is considered to be Lipschitz continuous if there exists $L>0$ such that
\begin{equation}\label{l-continuity}
\|F_n(x)-F_n(y)\|_2\leqslant L\|x-y\|_2,\kern0.2cm \forall\, x,y \in \mathbb{R}^n.
\end{equation}

% I THINK THIS SECTION SHOULD HAVE BEEN EXLUDED FROM PART 2 OF THE SURVEY. I HAVE COMMENTED OUT THE BELOW LINES AFTER WE SUBMITTED IT
% Many of the root finding methods described in Section \ref{rootfindingmethods} are guaranteed to converge to a solution when applied to SNEs that satisfy both the monotonicity and Lipchitz continuity conditions (the hybrid spectral methods introduced in \cite{aji2021} for example). 
% %For example, article \cite{aji2021} proposes two hybrid spectral methods and proves that they are guaranteed to converge when applied to SNEs which satisfy both the monotonicity and Lipchitz continuity conditions 

\subsection{Comparison to other surveys}
Other surveys discussing solution techniques for SNEs include \cite{Martinez1994} and \cite{gong2021nonlinear}. We have decided to conduct this comprehensive literature review because many new solution techniques for SNEs have been introduced since the publication of  \cite{Martinez1994} in 1994, and because the recent survey presented in \cite{gong2021nonlinear} focuses mainly on 
%almost exclusively discusses  
methods which first convert a SNE into an optimization problem, and then search for multiple solutions to the optimization problem using \textit{Intelligent Optimization Algorithms} (IOAs). The IOAs discussed in article \cite{gong2021nonlinear} are primarily metaheuristics for global optimization. Although the survey in article \cite{gong2021nonlinear} provides a very nice discussion of IOAs for solving SNEs reformulated as optimization problems, many of the IOAs they discuss are only introduced at a very high level, only eight IOAs were tested in their computational study, and the IOAs were evaluated on SNEs comprised of 20 equations or less. Also, article \cite{gong2021nonlinear} only briefly mentions methods which can be used to search for solutions to SNEs without transforming them into optimization problems.
%high level overview of many metaheuristics which solving the optimization reformulation of a SNE with metaheuristics,techniques based upon evolutionary algorithms, 

We would like to present a broader survey which covers in detail the large set of methods which can be used to solve a SNE without transforming it into an optimization problem (i.e.\ homotopy and symbolic computation methods). These methods were our main focus in part one of this survey. In part two, we expand upon article \cite{gong2021nonlinear} by introducing additional reformulation techniques and optimization algorithms which have been used to solve SNEs, and by discussing in much more detail many optimization algorithms which were only briefly introduced in article \cite{gong2021nonlinear}. This will allow us to appropriately set the stage for the comprehensive empirical study we will present in part three of this survey. Furthermore, we believe it is imperative to introduce the reader to a technique for determining the number of solutions to a SNE that exist within a bounded domain. Such techniques are of critical practical importance for those interested in finding all solutions to a SNE that exist within a domain of interest.

%many of these methods much care must be taken to a large array of methods in our survey, many of which are vastly different from one another, are vastly different that we discuss in our survey are very diverse cover a broad range of breadth and diversity of the methods that we discuss of 
%and we  highlight additional metaheuristics for global optimization which can be used to solve SNEs,
%as well as methods which can solve SNEs which do not require transforming a SNE into an optimization problem (i.e.\ homotopy methods).

% pays very little attention to techniques which do not require transforming a SNE into an optimization problem (i.e.\ homotopy methods), and only briefly discusses solving the global optimization reformulation of a SNE with techniques which are not based on evolutionary algorithms.

% other techniques which can be utfor solving the global optimization reformulation tdiscusses solving SNEs by transforming ution techniques for SNEs evolutionary algorithms. 

\subsection{Organization of this survey}
Before we discuss optimization methods which can be used to search for solutions to SNEs, we first introduce a method which can be used to determine the total number of solutions to a SNE that exist within a given bounded domain. If one can determine the number of solutions to a SNE which exist within a bounded domain of interest, one can confidently define an appropriate stopping criteria for algorithms which can be used to search for solutions. 
Next, 
% %After introducing a method for determining the number of solutions to a SNE within a given bounded domain, 
% we will introduce root finding methods which have been utilized in literature to search for solutions to a SNE without transforming a SNE into an optimization problem.
% %, First, we will discuss techniques which can be utilized to solve SNEs without transforming a SNE into an optimization problem. 
we will discuss the various ways that a SNE can be transformed into an optimization problem, and we will introduce techniques that can be utilized to search for solutions to the global optimization problem that arises when the most common reformulation is performed. 
% Afterwards, 
% %in Section \ref{additional-methods}
% we will introduce additional methods which have been used to attempt to solve SNEs, and
We will conclude our paper by highlighting promising areas for future research. %We will then conclude our paper 
%with a comparison of the solution techniques that we have discussed, 
% and highlight promising areas for future research.

% \section{Determination of the total number of solutions in a given bounded domain}\label{numsolutions}%\phantom{}\vspace*{0.1cm}
\section{Determining the number of solutions to a SNE in a bounded domain}\label{numsolutions}%\phantom{}\vspace*{0.1cm}

\noindent
The knowledge of all the solutions of a system of nonlinear equations and/or all the extrema of a function is of major importance in various fields.
The total number of the solutions of a system of nonlinear equations can be obtained 
by computing the topological degree.
Suppose that the function
$F_n=(f_1, f_2, \ldots,f_n)\colon {\mathcal D}_n \subset $ ${\mathbb R}^n$ $\to $ ${\mathbb R}^n$
is defined and is two times continuously differentiable in a bounded domain
${\mathcal D}_n$ of ${\mathbb R}^n$ with boundary $b({\mathcal D}_n)$. Suppose further that
the solutions of $F_n(x) = {\mathit \Theta}_n$
%(${\mathit \Theta }_n = (0,\ldots,0)$ denotes the origin of ${\mathbb R}^n$),
are not located on $b({\mathcal D}_n)$, and that they
are simple (that the determinant of the Jacobian of $F_n$ at these solutions is non-zero).
Then the {\em topological degree of $F_n$ at ${\mathit \Theta}_n$ relative to}
 ${\mathcal D}_n$ is denoted by
 ${\rm deg}[ F_n , {\mathcal D}_n,$ $ {{\mathit \Theta}_n }] $
and  can be defined by the following relation:
\begin{equation*}
 {\rm deg}[ F_n , {\mathcal D}_n , {{\mathit \Theta}_n} ]
=\displaystyle\sum_{{\scriptstyle x \in F_n^{-1} ( {{\mathit \Theta}_n} )} }
  {\rm sgn}\,{\rm det}\,J_{F_{n}} (x),\label{eq:deg}
\end{equation*}
where ${\rm det}\,J_{F_{n}} (x)$ denotes the determinant of the Jacobian matrix
and ${\rm sgn}$ defines the three-valued sign function.
The above definition can be generalized when $F_n$ is only continuous
{\cite{OrtegaR2000}}.

It is evident that, since  ${\rm deg}[ F_n , {\mathcal D}_n,$ $ {{\mathit \Theta}_n }] $
is equal to the number of simple solutions of $F_n(x) = {\mathit \Theta}_n$ which
give positive determinant of the Jacobian matrix, minus the number of simple solutions which give negative
determinant of the Jacobian matrix, then the total number $N^s$ of simple solutions of
$F_n(x) = {\mathit \Theta }_n$ can be obtained by the value of
${\rm deg}[ F_n , {\mathcal D}_n,$ $ {\mathit \Theta}_n] $ if all these solutions
have the same sign of the determinant of the Jacobian matrix. Thus, Picard considered
the following extensions of the function $F_n$ and 
the domain ${\mathcal D}_n$~\cite{Picard1892,Picard1922}:
%E. Picard, Sur le nombre des racines communes \`a plusieurs \'equations simultan\'ees, Journal de Math\'ematiques Pures et Appliqu\'ees Journ. de Math. Pure et Appl.($4^e$ s\'erie), 8 (1892), pp. 5-24.
%Émile Picard}, {\em Trait\'e d'analyse}, 3rd ed., chap. 4.7., Gauthier-Villars, Paris, 1922.
\begin{equation}\label{eq:pfn}
F_{n+1}=(f_1, f_2, \ldots,f_n,f_{n+1})\colon {\mathcal D}_{n+1} \subset
{\mathbb R}^{n+1}\to {\mathbb R}^{n+1}, 
\end{equation}
where\
$f_{n+1} = y\,\,{\rm det}\,J_{F_{n}}$,\,\, ${\mathbb R}^{n+1} : x_1, x_2,\ldots,x_n,y$,\, and
${\mathcal D}_{n+1}$ is the direct product of the domain ${\mathcal D}_{n}$
with an arbitrary interval of the real $y$-axis containing the point
$y=0$. Then the solutions of the following system of equations:
\begin{equation*}
\begin{array}{l}
f_{i}(x_1,x_2,\ldots,x_n) = 0,\quad i=1, 2,\ldots,n,\\[0.1cm]
y\,\,{\rm det}\,J_{F_{n}}(x_1,x_2,\ldots,x_n)=0,\label{eq:psys}
\end{array}
\end{equation*}
are the same simple solutions of $F_n(x) = {\mathit \Theta }_n$ provided that $y=0$.
Obviously, the determinant of the Jacobian matrix obtained for the function (\ref{eq:pfn})
is equal to $({\rm det}\,{J_{F_{n}}}(x))^2$ which is always positive.
Thus, the total number $N^s$ of the solutions of the system 
$F_n(x) = {\mathit \Theta }_n$ can be obtained by the following value of the topological degree:
\begin{equation*}
N^s = {\rm deg}[ F_{n+1}, {\mathcal D}_{n+1}, {\mathit \Theta }_{n+1} ].
\end{equation*}

For example, in the one dimensional case, using the above Picard's extensions it is proved that the total number of 
simple solutions $N^s $ of the equation $f(x)=0$, where $f\colon (a,b) \subset $ 
${\mathbb R}$ $\to $ ${\mathbb R}$ is twice continuously differentiable in a predetermined interval $(a,b)$, is given by the following relation~\cite{Picard1892,Picard1922}:
\begin{equation}\label{eq:NR}
{N^s} = - \frac{1}{\pi}\left[ \varepsilon
     \int_a^b \frac{f(x)\,f''(x)-{f'}^2(x)}{f^2(x) + {\varepsilon}^2 {f'}^2(x)}\, dx +
 \arctan\left(\frac{\varepsilon f'(b)}{f(b)} \right)
 - \arctan\left(\frac{\varepsilon f'(a)}{f(a)} \right) \right],
 \end{equation}
where $\varepsilon$ is a small positive constant. Note that ${N^s}$ was shown to be independent of
the value of $\varepsilon$. 
Also, the above approach can be applied for computing the number of multiple solutions.
Obviously, the total number  $N^e$ of the extrema of $f\in C^3$\,
i.e.\ $x \in (a,b)$ such that $f'(x)=0$ can be obtained using the above formula~(\ref{eq:NR}).
For details of the topological degree we refer the interested reader to the books~\cite{Cronin1964,Lloyd1978,OReganCC2006,OutereloR2009,OrtegaR2000,Sikorski2001}.
%Cronin, Jane [1964]. "Fixed Points and Topological Degree in Nonlinear Analysis,"
%American Mathematical Society, Providence, Rhode Island, 1964
%Sikorski, Krzysztof A.  Optimal Solution of Nonlinear Equations-Oxford University Press, New York (2001)
%Lloyd, Noel G. 1978 Degree theory, Cambridge University Press, New York, 1978.
%OrtegaR
%Outerelo, Enrique and Ruiz, Jes{\'u}s M., Mapping Degree Theory,
%Graduate Studies in Mathematics Volume 108, American Mathematical Society,Providence, Rhode Island, 2009
Details of the computation of the value of the topological degree and its usefulness as well as  
some applications and issues related to the number of zeros can be found for example 
in~\cite{BergaminBV2002,EmirisMV1999,KavvadiasV1996,Kearfott1979,MourrainPTV2006,MourrainVY2002,PlagianakosNV2001,PolymilisSSTV2003,Stenger1975,Stynes1979,Vrahatis1995,Vrahatis1988a,Vrahatis1989,VrahatisI1986,VrahatisSTB1993}. 

Article \cite{Franek2015} discusses the robust satisfiability of SNEs, and discusses utilizing the topological degree to determine the existence of robust solutions to a SNE.

\section{Transforming a SNE into an optimization problem}\label{reformulations}
Although many of the root finding methods described in part one of this survey are guaranteed to converge to a solution on certain classes of SNEs (such as SNEs that satisfy the conditions of monotonicity (cf. Eq.~(\ref{monotonicity})) and Lipschitz continuity (cf. Eq. (\ref{l-continuity}))), many of the root finding methods described in part one of this survey are incapable of reliably finding solutions to SNEs which do not satisfy such conditions. As a result, many approaches for solving general classes of SNEs first reformulate a SNE as an optimization problem, and then utilize optimization algorithms to attempt to find solutions. However, it is important to note that the resulting optimization problem may be multimodal and potentially nonsmooth. Therefore, global optimization techniques are often utilized to attempt to find solutions. For the interested reader, the books \cite{Boman1999,chinneck2008} discuss feasibility and infeasibility in optimization.

From the best of our knowledge, this section discusses the various techniques which have been presented in literature for reformulating a SNE as an optimization problem. In literature, techniques have been presented for transforming a SNE into a single objective global optimization problem (Section \ref{go-reformulation}), into a multiobjective optimization problem (Section \ref{mo-reformulation}), into a constrained optimization problem (Section \ref{co-reformulation}), and into a nonlinear programming problem (Section \ref{nl-reformulation}). Article \cite{gong2021nonlinear} provides a nice analysis and comparison of many of the reformulation techniques described in this section, especially in regards to the techniques they use to support efforts aimed at finding multiple solutions to a SNE. 

% \section{Transformation Techniques}
\subsection{Transforming a SNE into a single-objective global optimization problem}\label{go-reformulation}
A nonlinear system of equations (Eq.~(\ref{eq:1})) may be transformed into a single-objective global optimization problem by introducing the objective function
% $\varphi^0(x) = \sum_{k=1}^m|\varphi^{k}(x)|^p$ 
\begin{equation}\label{summation_transformation}
\varphi^0(x) = \alpha\sum_{i=1}^m|f_i(x)|^p
\end{equation}
% $\varphi^0(x) = \alpha\sum_{i=1}^m|f_i(x)|^p$ 
where $\alpha>0$ and $p>0$ (See Figure 2). Alternate formulations of $\varphi^0(x)$ have also been proposed in literature, for example, article \cite{ibrahim2019} proposes utilizing $\varphi^0(x) = \sqrt{\sum_{i=1}^m f_i^2(x)}$. In order to find solutions to a SNE (Eq.~(\ref{eq:1})), the following global optimization problem should be solved: \begin{equation}\label{eq:2}
\min_{x \in {\mathcal D}}\varphi^0(x).
\end{equation}
That is, to find the value of the {\em minimizer} $x$: 
\[
\arg \min_{x \in {\mathcal D}}\varphi^0(x),
\]
for which $\varphi^0(x)$ attains its {\em minimum} value. 

Although almost all papers which reformulate SNEs as optimization problems reformulate them as \textit{minimization} problems, SNEs can also be transformed into \textit{maximization} problems. Article \cite{Sidarto2015} proposes transforming SNEs into maximization problems by maximizing over an objective function of the form $\varphi(x) = {1}/({1+\sum_{i=1}^m |f_i(x)|})$. In this reformulation, solutions to a SNE are represented by points with an objective function value of 1. Article \cite{Sidarto2015} is the only work that we are aware of that reformulates SNEs as maximization problems. For the purposes of this paper, when we discuss reformulating a SNE as an optimization problem, we are referring to the standard case of reformulating a SNE as a minimization problem such as Eq. (\ref{eq:2}).

\begin{figure}[h]
  \centering
    \subfloat[\centering $\,\,\alpha=1,\, p=1$]{{\includegraphics[width=6.5cm]{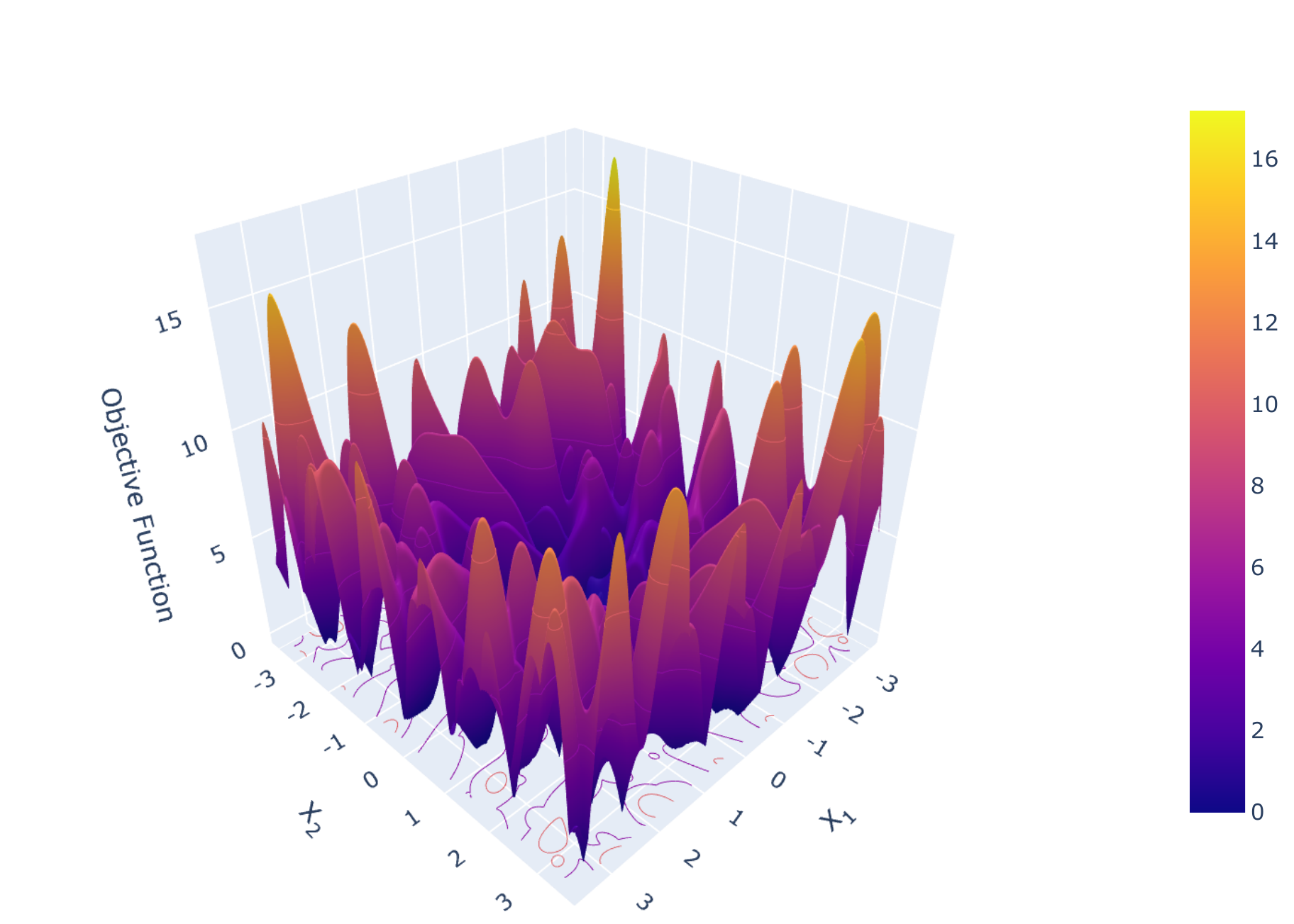} }}%
    \qquad
    \subfloat[\centering $\,\,\alpha=1,\, p=2$]{{\includegraphics[width=6.5cm]{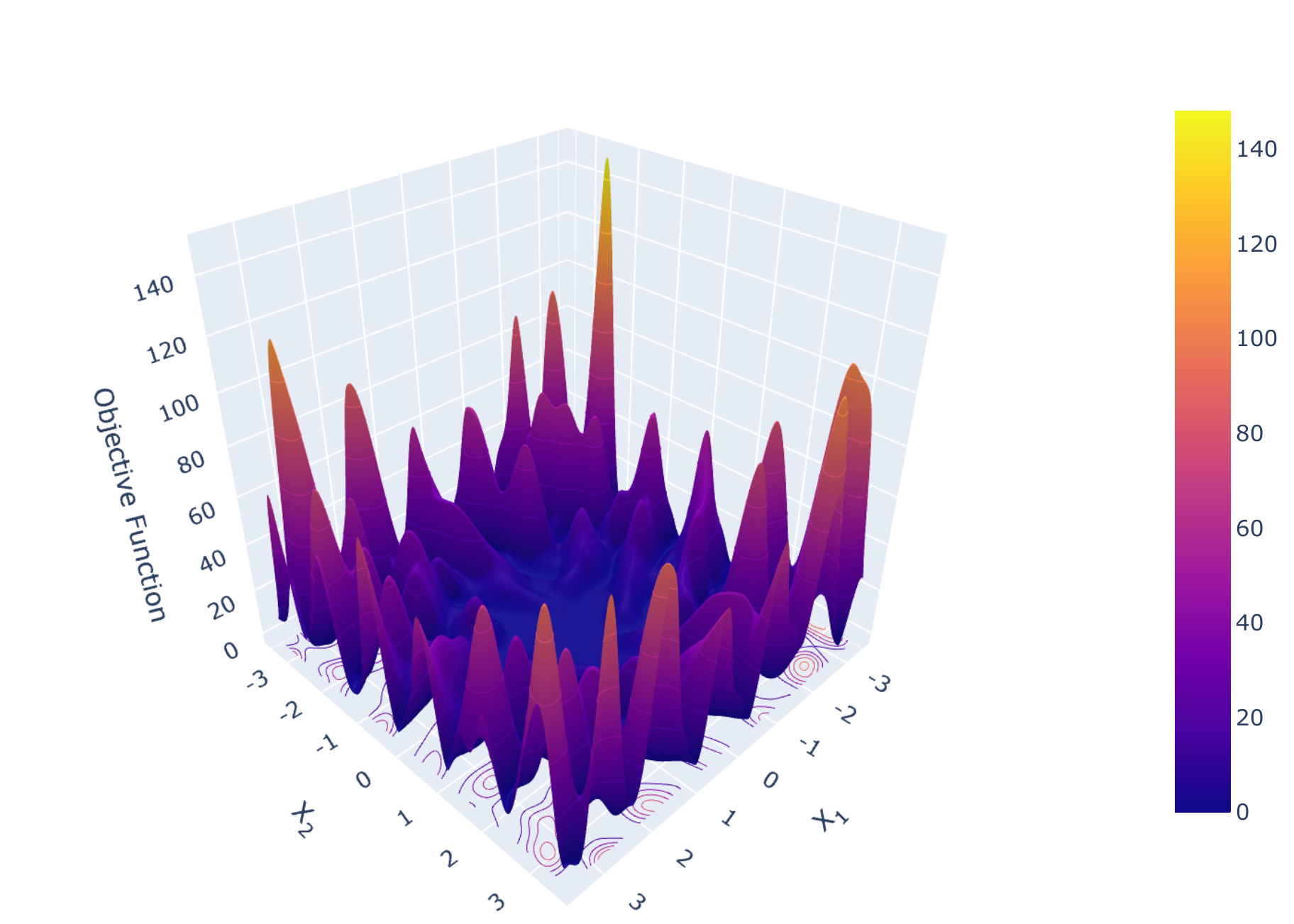} }}%
    \caption{The objective function surface 
    % $\varphi^0(x)=\sum_{k=1}^m |\varphi^{k}(x)|^p$ 
    $\varphi^0(x) = \alpha\sum_{i=1}^m|f_i(x)|^p$
    when $\alpha=1$, $p=1$ and $\alpha=1$, $p=2$ for the example SNE introduced in Eq.~(\ref{eq:00}) with two transcendental equations and two unknowns: \\
    $f_1(x_1,x_2) = x_1-x_1\sin(x_1+5x_2)-x_2\cos(5x_1-x_2)=0$; \kern0.1cm $f_2(x_1,x_2)=x_2-x_2\sin(5x_1-3x_2)+x_1\cos (3x_1+5x_2)=0$.}
%    $f_1(x,y)=x-x\sin(x+5y)-y\cos(5x-y)=0$; \kern0.3cm $f_2(x,y)=y-y\sin(5x-3y)+x\cos (3x+5y)=0$.}
%   \includegraphics[width=\linewidth]{SNE_example_desmos.png}
%   \caption{An example SNE with two equations and two unknowns: \\(Blue Equation): $x-x\sin(x+5y)-y\cos(5x-y)=0$; (Red Equation): $y-y\sin(5x-3y)+x\cos (3x+5y)=0$. Finding all of the points within a certain region which satisfy both equations is a challenging task. Graph generated by Desmos Graphing Calculator. (\url{https://www.desmos.com/calculator}).}
%   \Description{A woman and a girl in white dresses sit in an open car.}
\end{figure}
%

% \noindent
Since systems of nonlinear equations may have more than one solution, the reformulated optimization problem Eq. (\ref{eq:2}) may have more than one global minimizer, and each global minimizer with an objective function value of 0 (global minimum) will correspond to a solution to the SNE. Although the problem shown by Eq. (\ref{eq:2}) can be repeatedly solved to find many solutions to a SNE, depending on the solution technique used, the same solution(s) to a SNE may be found repeatedly during the search for new unfound solutions. To address this, a plethora of alternate reformulations have been utilized in literature which are specifically designed to help algorithms find further solutions to a SNE. These reformulations typically introduce a penalty (also referred to as a ``repulsion" \cite{Gong2020c} or ``polarization" \cite{HENDERSON2010551} technique) into the objective function to repel the search for new solutions away from solutions which have already been found. These reformulations are often of the form: 
%In addition to repeatedly solving There are multiple methods that transform SNE (eq. \ref{eq:1}) to an optimization problem.
% \subsection{Transformation Techniques for finding multiple solutions}
 %In order to find multiple solutions to a SNE, many papers utilize global optimization to attempt to solve modified objective functions of the form 
\begin{equation}\label{eq:modified}%todo: edit notatio
\varphi(x) = \varphi^0(x) + \varphi^{k}(x),
\end{equation}
where $ \varphi^{k}(x)$ creates a penalty region around the $k$ already found solutions (See Figure 3). Article \cite{pei2019} proposed to represent $\varphi^{k}(x)$ as: 
\begin{equation}\label{eq:modified1}
\varphi^{k}(x) = \sum_{j=0}^k{a\,\xi\left(\frac{\|x - x^j\|_2}{\rho}\right)},
\end{equation}
%%
%% here lets change \xi(x) to \xi of something else to avoid confusion
%%
where $x^j$ are already found solutions, $k$ is the number of solutions found so far, $a>0$ and $\rho>0$ are user defined parameters, and $\xi(\delta)$ is a continuous real function that is positive for $|\delta|<1$ and zero elsewhere. Article \cite{pei2019} proposes to utilize the following variants of $\xi(\delta)$:
\begin{itemize}
\setlength{\itemsep}{2pt}
    \item[(a)] $\xi(\delta) = 1 - |\delta|$ for $|\delta| < 1$, otherwise 0,
    \item[(b)] $\xi(\delta) = (1 - \delta^2)^2$ for $|\delta| < 1$, otherwise 0,
    \item[(c)] $\xi(\delta) = (1 + \cos(\pi \delta))/2$ for $|\delta| < 1$, otherwise 0,
     \item[(d)] $\xi(\delta) = \exp\bigl(-{\delta^2} / (1-\delta^2)\bigr)$ for $|\delta| < 1$, otherwise 0.
\end{itemize}
\begin{figure}[h]
  \centering
    \subfloat[\centering $\,\,\alpha=1,\, p=1,\, a=1,\, \rho=1$]
    % {{\includegraphics[width=6.5cm]{abs_w_penalty.pdf} }}%
    {{\includegraphics[width=6.5cm]{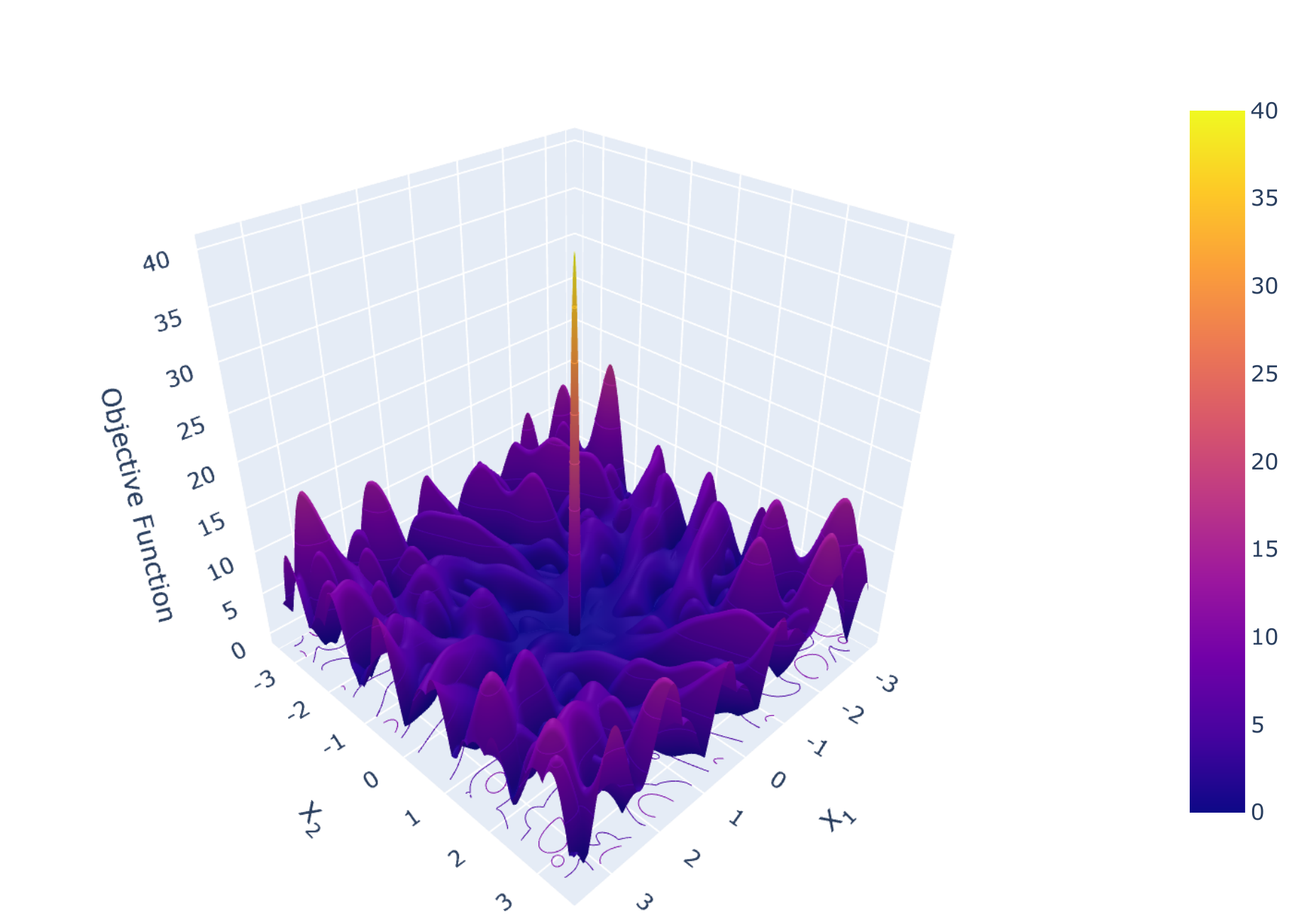} }}%
    \qquad
    \subfloat[\centering $\,\,\alpha=1,\, p=2,\, a=4,\, \rho=1$]
    % {{\includegraphics[width=6.5cm]{squared_w_penalty.pdf} }}%
    {{\includegraphics[width=6.5cm]{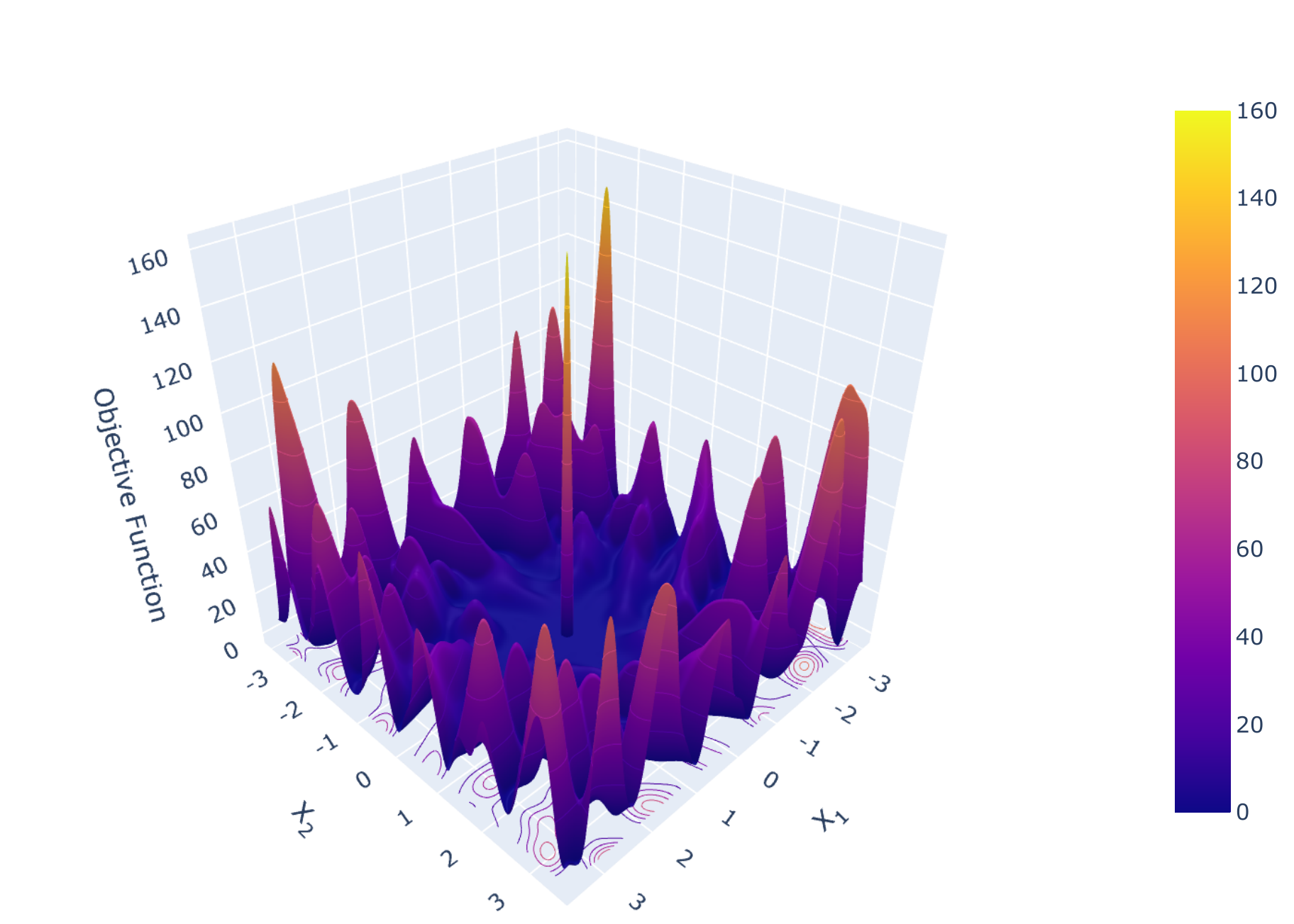} }}%
    \caption{The penalized objective function 
    % $\varphi(x)=\varphi^0(x)+\varphi^{k}(x)=\sum_{i=1}^m |\varphi^{i}(x)|^p+\sum_{j=0}^k{a\,\xi\left(\frac{\|x - x^j\|_2}{\rho}\right)}$ 
    $\varphi(x)=\varphi^0(x)+\varphi^{k}(x)=\alpha\sum_{i=1}^m|f_i(x)|^p+\sum_{j=1}^k{a\,\xi\left({\|x - x^j\|_2}/{\rho}\right)}$ 
    (where $\xi(\delta) = (2 - \delta^2)^2$ for $|\delta| < \sqrt{2}$ and $0$ otherwise) with a penalty $\varphi^{k}(x)$ for a single found solution ($k=1$) $x^1$ at $(0,0)$ when $\alpha=1, p=1, a=1, \rho=1$ and $\alpha=1, p=2, a=4, \rho=1$ for the example SNE introduced in Eq.~(\ref{eq:00}):  
        $f_1(x_1,x_2) = x_1-x_1\sin(x_1+5x_2)-x_2\cos(5x_1-x_2)=0$; \kern0.1cm $f_2(x_1,x_2)=x_2-x_2\sin(5x_1-3x_2)+x_1\cos (3x_1+5x_2)=0$.
    %$f_1(x,y)=x-x\sin(x+5y)-y\cos(5x-y)=0$; \kern0.2cm $f_2(x,y)=y-y\sin(5x-3y)+x\cos (3x+5y)=0$.
    Notice the sharply elevated region surrounding the point $(0,0)$ on both objective function surfaces.}
    %%The penalized objective function surface $F(x)=\varphi^0(x)+\varphi^{k}(x)=\sum_{k=1}^n |\varphi^{k}(x)|^p+\sum_{j=0}^k{a \xi\left(\frac{||x - x^j||_2}{\rho}\right)}$ with a penalty $\varphi^{k}(x)$ for a found solution at $(0,0)$ when $p=1$ and $p=2$ for the example SNE introduced in Eq.~{eq:00}: $f_1(x,y)=x-x\sin(x+5y)-y\cos(5x-y)=0$; $f_2(x,y)=y-y\sin(5x-3y)+x\cos (3x+5y)=0$. In figure (a) on the left, $\xi(x) = (2 - x^2)^2$ for $|x| < 1$ and $0$ otherwise. In figure (b) on the right, $\xi(x) = 4(2 - x^2)^2$ for $|x| < 1$ and $0$ otherwise. Notice the sharply elevated regions surrounding the point $(0,0)$ on both surfaces.
    
%   \includegraphics[width=\linewidth]{SNE_example_desmos.png}
%   \caption{An example SNE with two equations and two unknowns: \\(Blue Eq.~): $x-x\sin(x+5y)-y\cos(5x-y)=0$; (Red Eq.~): $y-y\sin(5x-3y)+x\cos (3x+5y)=0$. Finding all of the points within a certain region which satisfy both equations is a challenging task. Graph generated by Desmos Graphing Calculator. (\url{https://www.desmos.com/calculator}).}
%   \Description{A woman and a girl in white dresses sit in an open car.}
\end{figure}
Notice that if there are no solutions found yet, then Eq.~(\ref{eq:modified}) reduces to the original optimization problem Eq. (\ref{eq:2}). 
A similar approach is suggested in the papers \cite{hirsch2009, Silva2014} where 
\begin{equation}\label{eq:modified2}
\varphi^{k}(x) = \beta \sum_{j=0}^k {\exp\bigl(-\|x-x^j\|_2)\, \chi(\|x-x^j\|_2\bigr)}.
\end{equation}
Here, $\beta$ is a large constant, $\varrho$ is a small constant, $\chi(\delta) = 1$ when $\delta < \varrho$ and otherwise $\chi(\delta) = 0$. 
%%%%%%%%%%%%%%%%%%%%%%%%%%%%%%%%%
%% Note that many of the reformulations below
%% were copied almost exactly from https://ieeexplore.ieee.org/stamp/stamp.jsp?arnumber=9426462
%% so we may need to add some commentary, adjust notation, or something else unique prior to submission
%%%%%%%%%%%%%%%%%%%%%%%%%%%%%%%%%
%\noindent
For the rest of the reformulations introduced in Section \ref{go-reformulation}, we assume that at least one solution has been found so far ($k\geqslant 1)$.

Article \cite{HENDERSON2010551} proposes minimizing an objective function of the form 
\begin{equation}\label{eq:modified5}
\varphi(x)=\dfrac{\varphi^0(x)}{\Pi_{j=1}^k \arctan \|x-x^j\|_2},    
\end{equation}
while article \cite{pourjafari2012} proposes minimizing an objective function of the form 
\begin{equation}\label{eq:modified51}
\varphi(x)= \bigl(\varphi^0(x)+\varepsilon\bigr)\, \Pi_{j=1}^k \bigl|\coth(\alpha \|x-x^j\|_2)\bigr|, 
\end{equation}
where $\varepsilon>0$ is a small user defined constant and $\alpha\geqslant 1$ is a user defined parameter which is utilized to adjust the radius of the penalty region. 

% \noindent
Article \cite{Ramadas2014} proposes minimizing an objective function of the form 
\begin{equation}\label{eq:modified6}
\varphi(x)=\varphi^0(x)\, \Pi_{j=1}^k \xi(x, x^j, \alpha, \rho), 
\end{equation}
where 
\begin{align*}
  \xi(x, x^j, \alpha, \rho)=
   |\,\text{erf}(\alpha \|x-x^j\|_2)|^{-1}\,\, \text{ if }\,\, \|x-x^j\|_2\leqslant \rho, \text{ and }\, 0 \text{ otherwise}.
\end{align*}
Here, $\rho>0$ and $\alpha>0$ are user defined parameters which adjust the radius of the penalty region and the magnitude of the penalty respectively, and $\text{erf}(x)=\frac{2}{\sqrt{\pi}}\int^x_0 \exp (-t^2)dt$. 
%The authors suggested setting $\rho$ to be equal to one tenth of the width of the smallest dimension of the hyperrectangle defining the feasible region, i.e. that 
%$\rho=0.1\,\min_{i=1,\ldots,n}|a_i-b_i|$. 

Article \cite{Liao2020} proposes a similar reformulation to Eq. (\ref{eq:modified6}) with a dynamic repulsion radius that is updated within a evolutionary algorithm,
and
article \cite{Ramadas2015} proposes minimizing an objective function of the form: 
\begin{equation}\label{eq:modified7}
\varphi(x)=\varphi^0(x)+\sum_{j=1}^k \xi(x, x^j, \alpha, \rho), 
\end{equation}
where 
\begin{align*}
  \xi(x, x^j, \alpha, \rho)=
   \alpha\bigl(1-\text{erf}(\|x-x^j\|_2)\bigr)\, \text{ if }\, \|x-x^j\|_2\leqslant \rho, 
   \text{ and }\, 0\text{ otherwise},
\end{align*}
where $\rho>0$ and $\alpha>0$ are user defined parameters which adjust the radius of the penalty region and the magnitude of the penalty respectively, and where $\text{erf}(x)=\frac{2}{\sqrt{\pi}}\int^x_0 \exp (-t^2)dt$.

Other single objective reformulations have been utilized for particular applications. For example, articles \cite{henderson2004,FREITAS2004} propose minimizing over an objective function of the form: 
\begin{equation}\label{eq:modified3}
% \varphi(x)=\varphi^0(x)\, \dfrac{\Pi_{i=1}^n x_i^{\alpha\,\!k}}{\Pi_{j=1}^k \Pi_{i=1}^n \bigl|x_i-x_i^j \bigr|^\alpha},   
\varphi(x)=\varphi^0(x)\,  \Pi_{j=1}^k \left( \frac{\Pi_{i=1}^n ((b_i)^{k \alpha})}{\Pi_{i=1}^n |x_i-x^k_i|^\alpha} \right),
\end{equation}
\noindent
where $a\leqslant x \leqslant b$, $x^j$ for $ j=1,2,\ldots,k$ are already found solutions ($k$ is the number of solutions found so far), $x_i$ is the $i-$th coordinate of the solution $x$, and $\alpha>0$ is a user defined parameter. Article \cite{FREITAS2004} also proposes a slightly different objective function, they propose minimizing over an objective function of the form
\begin{equation}\label{eq:modified4}
%   \varphi(x)=\varphi^0(x)\, \Pi_{j=1}^k \exp\left(\dfrac{\Pi_{i=1}^n x^k}{\Pi_{i=1}^n \bigl|x^k-x^k^j \bigr|}\right).
\varphi(x)=\varphi^0(x)\, \Pi_{j=1}^k \exp\left( \frac{\Pi_{i=1}^n ((b_i)^{k \alpha})}{\Pi_{i=1}^n |x_i-x^k_i|^\alpha} \right).
\end{equation}
It is important to note that for the reformulations in Eq. (\ref{eq:modified3})
and Eq. (\ref{eq:modified4}), if any of the coordinates of a new solution are the same as that of a previous found solution, the objective function value will be infinity. By using this reformulation, one can ensure that for each new solution that is found, every dimension will be different from that of all previously found solutions.

\subsection{Transforming a SNE into a single-objective constrained optimization problem}\label{co-reformulation}
In addition to the non-constrained reformulations we have described so far, constrained reformulations include: 
\begin{equation}
    \min \varphi(x) \,\, \text{ for }\,\,  \varphi(x) = \sum_{i=1}^m f_i(x),
\end{equation}
subject to 
\[
f_i(x)\geqslant 0,\kern0.3cm \forall\, i = 1, 2,\ldots, m,
\]
as presented in \cite{kuri2003}. Alternatively, article \cite{Pourrajabian2013} first partitioned the SNE into two sets of equations $S_1=\{1,2,\ldots,m\}$ and $S_2=\{1,2,\ldots,m\}/S_1$. Based on these two sets of equations, article \cite{Pourrajabian2013} then utilized the reformulation: 
\begin{equation}\label{constrained}
    \min \varphi(x) \,\, \text{ for }\,\,  \varphi(x) = \sum_{i\in S_1} f_i^2(x),
\end{equation}
subject to 
\[
f_j(x)= 0,\kern0.3cm \forall\, j \in S_2.
\]
%Since all solutions to a SNE have an objective function value of zero when using reformulation (\ref{constrained}), some may consider the constraints in reformulation (\ref{constrained}) to be unnecessary or redundant.
Constrained optimization based transformation techniques such as those presented above require a constraint handling method to solve. %When utilizing reformulating SNEs Since it is clear that all solutions to a SNE have an objective function value of zero, it 

\subsection{Transforming a SNE into a multiobjective optimization problem}\label{mo-reformulation}
Instead of transforming a SNE into a single-objective optimization problem, a number of research papers (see for example \cite{Grosan2008,6849952}) 
%cite{Grosan2008}, \cite{6849952} 
suggest transforming a SNE 
(Eq.~(\ref{eq:1})) into a multiobjective optimization problem of the form
\begin{equation}\label{eq:multiobj}
  \left\{\begin{aligned}
   &\text{min } |f_1(x_1,x_2,\ldots,x_n)|,\\[0.2cm]
   &\text{min } |f_2(x_1,x_2,\ldots,x_n)|,\\
   & \kern1.7cm \vdots \\
   % \vdotswithin{=} \\ %how to move the dots?
   &\text{min } |f_m(x_1,x_2,\ldots,x_n)|,
\end{aligned} \right.
\end{equation}
which can be solved via multiobjective optimization methods.
Alternatively, instead of having an objective function for each equation, article \cite{Naidu2018} randomly splits the $m$ equations into two subsets $S_1$ and $S_2$ such that $S_1 \cap S_2 =\emptyset$ and $S_1 \cup S_2 $ is equivalent to the SNE (\ref{eq:1}). In terms of these two sets, article \cite{Naidu2018} then transforms a SNE into a biobjective optimization problem of the form: 
\begin{equation}\label{eq:multiobj5}
  \left\{\begin{aligned}
   &\min \varphi^1(x) \,\, \text{ for }\,\, \varphi^1(x) =\sum_{i=1}^{|S_1|}|f_{S_1(i)}(x)|, \text{ s.t. } f_{S_1(i)}(x)=0, \\[0.2cm]
   &\min \varphi^2(x) \,\, \text{ for }\,\, \varphi^2(x) =\sum_{i=1}^{|S_2|}|f_{S_2(i)}(x)|, \text{ s.t. } f_{S_2(i)}(x)=0, \\
\end{aligned} \right.
\end{equation}
or 
\begin{equation}\label{eq:multiobj6}
  \left\{\begin{aligned}
   &\min \varphi^1(x) \,\, \text{ for }\,\, \varphi^1(x) = \sum_{i=1}^{|S_1|}|f_{S_1(i)}(x)|, \text{ s.t. } f_{S_1(i)}(x)\geqslant 0, \\[0.2cm]
   &\min \varphi^2(x) \,\, \text{ for }\,\, \varphi^2(x) = \sum_{i=1}^{|S_2|}|f_{S_2(i)}(x)|, \text{ s.t. } f_{S_2(i)}(x)\geqslant 0.
\end{aligned} \right.
\end{equation}

% \noindent
Article \cite{6849952} reformulates a SNE into a biobjective optimization problem of the form: 
\begin{equation}\label{eq:multiobj2}
  \left\{\begin{aligned}
   &\min \varphi^1(x) \,\, \text{ for }\,\, \varphi^1(x) = x_1 + \sum_{i=1}^m |f_i(x)|,\\[0.2cm]
   &\min \varphi^2(x) \,\, \text{ for }\,\, \varphi^2(x) = 1-x_1+m\,\max |f_i(x)|,\\
\end{aligned} \right.
\end{equation}
where $i = 1, 2, \ldots, m$, and $x_1$ represents the first decision variable of $x$. In this reformulation, $x_1$ is utilized to ensure that the two objective functions conflict. Note that when using this reformulation, it may be difficult to detect multiple solutions that have the same value of $x_1$ \cite{gong2021nonlinear}. 
Expanding upon this approach, article \cite{Gong2017} presented a biobjective reformulation of the form: 
\begin{equation}\label{eq:multiobj3}
  \left\{\begin{aligned}
   &\min \varphi^1(x) \,\, \text{ for }\,\, \varphi^1(x) = \frac{\sum_{i=1}^n w_i\, x_i}{\sum_{i=1}^n w_i} + \sum_{j=1}^m |f_j(x)|,\\[0.2cm]
   &\min \varphi^2(x) \,\, \text{ for }\,\, \varphi^2(x) = 1-\frac{\sum_{i=1}^n w_i\, x_i}{\sum_{i=1}^n w_i} + \sum_{j=1}^m |f_j(x)|,\\
\end{aligned} \right.
\end{equation}
where $i=1, 2, \ldots,n$, $w_i$ is a randomly generated weight for each variable such that $\sum_{i=1}^n w_i=1$, $n$ is the number of decision variables in $x$, and $m$ is the number of equations. By utilizing this reformulation, the risk of not being able to identify two solutions with the same value of $x_1$ inherent in reformulation (\ref{eq:multiobj2}) is decreased. 

Also expanding upon reformulation (\ref{eq:multiobj2}), article \cite{Qin2015} 
%expands upon reformulation (\ref{eq:multiobj2}) by 
transforms a SNE into a multiobjective optimization problem with $n+1$ objectives where $n$ is the number of decision variables in the SNE. The reformulated multiobjective optimization problem is of the form: 
\begin{equation}\label{eq:multiobj9}
  \left\{\begin{aligned}
   &\min \varphi^1(x) \,\, \text{ for }\,\, \varphi^1(x) = \frac{x_1}{n}+\frac{x_2}{n-1}+\cdots+\frac{x_{n-1}}{2}+\frac{x_n}{1}+C\, \sum_{i=1}^m|f_i(x)| \, \ln(n+2),\\
   &\min \varphi^2(x) \,\, \text{ for }\,\, \varphi^2(x) = \frac{x_1}{n}+\frac{x_2}{n-1}+\cdots+\frac{x_{n-1}}{2}+(1-x_{n})+C\, \sum_{i=1}^m|f_i(x)| \, \ln(n+1),\\
   &\min \varphi^3(x) \,\, \text{ for }\,\, \varphi^3(x) = \frac{x_1}{n}+\frac{x_2}{n-1}+\cdots+\frac{x_{n-2}}{3}+(1-x_{n-1})+C\, \sum_{i=1}^m|f_i(x)| \, \ln(n),\\
            &\kern5.62cm \vdots\\
   &\min \varphi^n(x) \,\, \text{ for }\,\, \varphi^n(x) = \frac{x_1}{n}+(1-x_2)+C\, \sum_{i=1}^m|f_i(x)| \, \ln(3),\\
   &\min \varphi^{n+1}(x) \,\, \text{ for }\,\, \varphi^{n+1}(x) =(1-x_1)+C\, \sum_{i=1}^m|f_i(x)| \, \ln(2).
\end{aligned} \right.
\end{equation}

In article \cite{Gao2021}, a SNE is transformed into a biobjective optimization problem of the form: 
\begin{equation}\label{eq:multiobj7}
  \left\{\begin{aligned}
   &\min \varphi^1(x) \,\, \text{ for }\,\, \varphi^1(x) = \sum_{i=1}^m [f_i(x)]^2,\\[0.2cm]
   &\min \varphi^2(x) \,\, \text{ for }\,\, \varphi^2(x) = -\frac{1}{S_n} {\sum_{j=1}^{S_n} K(x,x^j)},\\
\end{aligned} \right.
\end{equation}
where $S_n$ is the size of the population utilized in the article's proposed evolutionary algorithm, $x^j$ for $j=1,\ldots,S_n$ are the elements of the current population, $K(x,x^j) = \exp \bigl(-\|x-x^j\|_2 / {(2 \sigma^2)} \bigr)$ is the Gaussian kernel of $x$ and $x^j$, and $\sigma>0$ is the kernel radius. In \cite{Gao2021}, the authors set $\sigma=2$. Here, the first objective focuses on exploitation while the second focuses on exploration. There is no guarantee that the two objectives $\varphi^1(x)$ and $\varphi^2(x)$ will conflict. 
%\begin{equation}\label{eq:multiobj8}
%  \left\{\begin{aligned}
%   &\min \varphi^1(x) \,\, \text{ for }\,\, \varphi^1(x) = \sum_{j=1}^m f^2_j(x),\\[0.2cm]
%   &\min \varphi^2(x) \,\, \text{ for }\,\, \varphi^2(x) = -\frac{1}{S_n} {\sum_{j=1}^{S_n} K(x,x_j)}.\\
%\end{aligned} \right.
%\end{equation}

Article \cite{JI2021204} reformulates a SNE as a dynamic tri-objective optimization problem, and introduces a self-adaptive ranking multi-objective differential evolution algorithm which is utilized to solve the dynamic tri-objective optimization problem. The proposed tri-objective optimization problem is of the form

\begin{equation}\label{eq:multiobj8}
  \left\{\begin{aligned}
   &\min \varphi^1(x) \,\, \text{ for }\,\, \varphi^1(x) = \frac{\sum_{j=1}^{S_n} \|x-r^j\|_2}{\max_{t=1,\ldots,P_s}\sum_{j=1}^{S_n} \|x^t-r^j\|_2}+\frac{9 n g^3}{\gamma^3}\left( \sum_{i=1}^m |f_i(x)| + \max_{i=1,\ldots,m}|f_i(x)|\right),\\[0.3cm]
   &\min \varphi^2(x) \,\, \text{ for }\,\, \varphi^2(x) = 1-\frac{\sum_{j=1}^{S_n} \|x-r^j\|_2}{\max_{t=1,\ldots,P_s}\sum_{j=1}^{S_n} \|x^t-r^j\|_2}+\frac{9 n g^3}{\gamma^3}\left( \sum_{i=1}^m |f_i(x)| + \max_{i=1,\ldots,m}|f_i(x)|\right),\\[0.3cm]
    &\min \varphi^3(x) \,\, \text{ for }\,\, \varphi^3(x) =\sum_{j=1}^{S_n}\max \left(1-\frac{\|x-x^j\|}{\frac{\sqrt{n}}{\sqrt[n]{S_n}}(1-\frac{g}{\gamma})},0\right)+\frac{9 n g^3}{\gamma^3}\left( \sum_{i=1}^m |f_i(x)| + \max_{i=1,\ldots,m}|f_i(x)|\right),\\
\end{aligned} \right.
\end{equation}
where $S_n$ is the population size utilized in the proposed evolutionary algorithm,  $r^j$ for $j=1,2,\ldots,S_n$ are randomly generated reference points at the initial stage, $x^t$ for $ t=1, 2, \ldots, S_n$ are the current points in the population, $n$ is the dimension of the solution vector $x \in \mathbb{R}^n$, $g$ is the index of the current evolutionary generation, and $\gamma$ is the maximum number of generations over which the evolutionary algorithm is allowed to evolve. 

Since the proposed bi-objective and tri-objective optimization problems introduced in Eq. (\ref{eq:multiobj7}) and Eq. (\ref{eq:multiobj8}) contain parameters which depend on the configuration of algorithms proposed in the original papers, we refer the interested reader to the source articles \cite{JI2021204,Gao2021} for additional details. %self-adaptive ranking multi-objective differential evolution algorithm which they propose, we refer the interested reader to the original article 

% Article \cite{ibrahim2019} proposes utilizing an objective function of the form: 
% \[
% \min \varphi(x)  \,\, \text{ for }\,\,  \varphi(x) = \sqrt{\sum_{i=1}^m f_i^2(x)}.
% \]

%\cite{naidu2016solving} appears to use the typical multi-objective reformulation where each equation is an objective.
%JI2021204 contains new method of transformation
% \noindent
% \cite{Gao2020, Gao2021}

\subsection{Transforming a SNE into a nonlinear programming problem}\label{nl-reformulation}
SNEs can also be explicitly reformulated as nonlinear programming problems. 
Article \cite{Mousa2008} reformulates a square SNE (such that $m=n$) into a nonlinear programming problem of the form: 
\begin{equation}
    \min\,\bigl\|\varepsilon_i-\varepsilon_i^*\big\|_\infty,
\end{equation}
subject to 
\begin{equation}
  \left\{\begin{aligned}
  \ \ & \\[-0.45cm]
   \ \ &\bigl|f_i(x)\bigr|-\varepsilon_i \leqslant 0,\\[0.3cm]
  \  \ &\varepsilon_i \geqslant 0,\\[0.05cm] 
\end{aligned} \right.
\end{equation}
where $i=1, 2, \ldots, n$ and $\varepsilon^*\in \mathbb{R}^n$ is the precision of the desired solution (meaning that $\varepsilon_i^*=0$ corresponds to a true solution to a SNE).

\section{Optimization techniques to find solutions to a SNE}\label{optimizationmethods}
By formulating SNEs of interest as optimization problems, many techniques traditionally utilized for finding one or more minima of a function can be applied to attempt to find solutions to SNEs. There are many well known optimization algorithms which can be used to attempt to find one or more minima of an objective function, and the performance of these methods can vary considerably depending on the characteristics of the objective function. However, some optimization algorithms can only be applied to objective functions with certain characteristics (of differentiability class $C^2$ for example). Since some of the reformulations described in Section \ref{reformulations} can result in non-smooth and multimodal objective function surfaces, it is important that the characteristics of the reformulated objective function are considered when a solution technique is selected. %If the characteristics of the reformulated objective function are unknown, a solution technique capable of handling non-smooth and multimodal objective functions should be considered.

In cases where the characteristics of the reformulated objective function are difficult to determine, one should consider utilizing an optimization algorithm which can be applied to non-smooth and multimodal objective functions. If one can assume that the reformulated objective function is unimodal and smooth (neither of which is true in the general case for SNEs), one could simply use a local search method to attempt to find a solution to a SNE via minimization. If one can only assume that the reformulated objective function is unimodal, then a gradient-free local search technique should be used. If one can not assume that the objective function is unimodal, then a local search method alone is unlikely to be sufficient for finding a solution to a SNE. In the multimodal case, one should attempt to find solutions to a SNE by using an algorithm for global optimization, such as those described in Section \ref{metaheuristics}. 
%  (which is highly unlikely in the general case for NSEs) that   of interest by such as the derivative-free methods describ with
 %Some of the methods described in {\em Newton's method} and the {\em Quasi-Newton methods} described in Section \ref{rootfinding}. 
%including many of the methods described in section  and the following sections discuss a few of these methods which have been specifically utilized to find solutions to SNEs as described in the referenced literature.
% \subsection{Newton's method for optimization}
Although the scope of our paper is focused on discussing techniques which have been directly utilized in literature to search for solutions to SNEs, there are many optimization algorithms which have yet to be utilized to search for solutions to SNEs. In addition to learning about the techniques explicitly described in this paper, we suggest that the interested reader learn more about the taxonomy of global optimization algorithms by reviewing articles such as \cite{Stork2020}. 

In this paper, we have classified the optimization algorithms which have been utilized to search for solutions to SNEs into two categories: (1) \textit{local search methods}, and (2) \textit{global search methods}. Section \ref{localsearch} describes a few of the many local search techniques which can be used to attempt to find a point $x'$ sufficiently close to a critical point on the objective function surface. Therefore, a solution to a SNE can be found if a local search algorithm terminates at a point $x'$ which has an objective function value sufficiently close to zero. Since the reformulated objective function of an arbitrary SNE may be highly multimodal, local search techniques are often utilized within \textit{hybrid metaheuristics} and \textit{memetic algorithms} to find solutions to SNEs via global optimization. Section \ref{metaheuristics} focuses on introducing metaheuristics which have been utilized to find solutions to SNEs via global optimization. The book \cite{pardalos2018} provides a nice introduction to heuristics for global optimization, and the books \cite{souravlias2019parallel,souravlias2021algorithm} discuss strategies for developing portfolios of such optimization algorithms. Unless otherwise noted, in Section \ref{optimizationmethods} we will describe techniques which can be used to solve SNEs reformulated into single objective non-constrained global optimization problems such as Eq.~(\ref{eq:2}).

\subsection{Local search methods}\label{localsearch}

Since Newton's method is capable of finding the roots of a differentiable function, Newton's method can be applied to the gradient $\nabla \varphi(x)$  of a twice differentiable function $\varphi(x)$ to find critical points. Therefore, if an objective function such as Eq.~(\ref{eq:modified}) is twice differentiable, Newton's method can be applied from an arbitrary starting point $x^0 \in \mathcal{D} \subset \mathbb{R}^n$ to find a nearby local minimum or saddle point on the objective function surface. Once Newton's method terminates at a point $x' \in \mathcal{D} \subset \mathbb{R}^n$, one can then check whether or not the objective function value at $x'$ is sufficiently close to zero to determine whether or not $x'$ is a solution to the NSE of interest (since every solution to a SNE corresponds to a point with an objective function value of zero). 

% \subsection{Gradient and conjugate gradient methods}
% section which has \nabla in the update formula
%  If $\varphi^0(x)$ is differentiable, then {\em nonlinear gradient methods} and {\em nonlinear conjugate gradient methods} %{\em gradient methods}\/ and {\em conjugate gradient methods}
% may be used to attempt to solve it. {\em Gradient descent methods} are based on the observation that the function $\varphi(x)$ decreases the most in the direction opposite to its gradient $\nabla \varphi(x)$. Hence, starting from an initial solution $x^0\in \mathcal{D} \subset \mathbb{R}^n$, the {\em gradient descent method} iteratively finds a sequence of new solutions by calculating $x^{i+1} = x^i - \alpha \nabla \varphi(x^i)$. There are multiple methods for determining the step size $\alpha$. One well known method for determining the step size $\alpha$ is the {\em Barzilai-Borwein method} \cite{barzilai1988}. Article \cite{fletcher2005barzilai} provides a nice introduction to the {\em Barzilai-Borwein method} and compares the {\em Barzilai-Borwein method} against many alternative methods.
 If the objective function is differentiable, then {\em nonlinear gradient methods} and {\em nonlinear conjugate gradient methods} %{\em gradient methods}\/ and {\em conjugate gradient methods}
may be used to attempt to find one if its critical points. Gradient descent methods are based on the observation that the function $\varphi(x)$ decreases the most in the direction opposite to its gradient $\nabla \varphi(x)$. Hence, starting from an initial solution $x^0\in \mathcal{D} \subset \mathbb{R}^n$, the gradient descent method iteratively finds a sequence of new solutions by calculating $x^{i+1} = x^i - \alpha^i s^i$ where $s^i=\nabla \varphi(x^i)$ is the search direction and $\alpha^i$ is the step size. There are multiple methods for determining the step size $\alpha^i$, including the well known {\em Barzilai-Borwein method} \cite{barzilai1988}. Article \cite{fletcher2005barzilai} provides a nice introduction to the Barzilai-Borwein method and compares the Barzilai-Borwein method against alternative methods. Overview of gradient descent optimization methods may be found in \cite{DBLP:journals/corr/Ruder16}. Popular extension of the gradient descent method is called \textit{stochastic gradient descent}, that can be efficient when the function $\varphi(x)$ can be represented as a sum of large number of functions $\varphi(x) = \sum_i^K\varphi_i(x)$. Then, the gradient is estimated on a random subset of $\varphi_i(x)$. Widely used stochastic gradient descent algorithms are ADAM \cite{DBLP:journals/corr/KingmaB14}, AdaGrad\cite{10.5555/1953048.2021068}, and others.

% Alexander's original section
% {\em Conjugate gradient methods}\/ differ from the {\em gradient descent method} by introducing additional steps based on conjugate gradients into calculating the search direction $s^i$. Multiple methods for updating the conjugate gradient search direction were proposed in the literature. These methods include the Fletcher–Reeves \cite{fletcher1964}, Polak–Ribière \cite{polak1969note}, Hestenes-Stiefel \cite{hestenes1952methods}, and \cite{dai1999nonlinear} methods among others.
% Article \cite{waziri2020} applies the {\em conjugate gradient method} to solve SNEs, and it suggests combining the Fletcher–Reeves and Polak–Ribière methods with the Quasi-Newton update. The paper first reformulates a SNE into an unconstrained optimization problem of the form XXXX, and then searches for a solution. Article \cite{waziri2020} numerically evaluates the performance of the proposed method on 20 test problems with up to 10k dimensions, and most of the problems were solved by the proposed method in under 3 seconds and with less than 30 iterations of the algorithm. Related approaches, also based on the same reformulation of SNEs, are proposed in \cite{dauda2016,dauda2020}. 
% Will's revised section
Conjugate gradient methods differ from gradient descent methods by using the conjugate search direction $s^i=-\nabla \varphi(x^i)+\beta^is^{i-1}$. Multiple methods for calculating the parameter $\beta^i$ of the conjugate search direction have been proposed in the literature. Commonly utilized methods for calculating $\beta^i$ were introduced by Fletcher–Reeves \cite{fletcher1964}, Polak–Ribière \cite{polak1969note}, Hestenes-Stiefel \cite{hestenes1952methods}, and Dai-Yuan \cite{dai1999nonlinear} among others.
Many variants based on conjugate gradient methods have been proposed in literature. For example, articles \cite{aji2020modified, Mahdi2021,Ahookhosh2013,mahdi2020} propose new methods which combine hyperplane projection techniques with the conjugate gradient method to solve systems of monotone nonlinear equations (SNEs which satisfy the inequality in Eq. (\ref{monotonicity})).
Article \cite{waziri2020} applies the conjugate gradient method to solve SNEs, and it suggests combining the Fletcher–Reeves and Polak–Ribière methods with the Quasi-Newton update. The paper first reformulates a SNE into a unconstrained single objective optimization problem, and then searches for a solution. Article \cite{waziri2020} numerically evaluates the performance of the proposed method on 20 test problems with up to 10k dimensions, and most of the problems were solved by the proposed method in under 3 seconds and with less than 30 iterations of the algorithm when coded in MATLAB and run on a computer with a 2.13 GHz CPU processor and 2GB of RAM. Additional works which discuss using conjugate gradient based methods for solving SNEs reformulated as single objective optimization problems include \cite{dauda2016,dauda2020}. 

% {\em Spectral gradient methods} for minimization are another class of local search methods which have been utilized to find solutions to NSEs. Spectral gradient methods determine the search direction $s^i$ by in a variety of ways, do not require knowledge of the Jacobian matrix, and hence can be used and can be shown to converge under some specific conditions \cite{aji2021}. Other papers discussing utilizing spectral methods to find solutions to SNEs include \cite{LaCruz2006}.

In addition to Quasi-Newton methods, other commonly utilized derivative-free direct search methods worth mentioning include the {\em Rosenbrock search method} \cite{rosenbrock1960}, the {\em Hooke and Jeeves pattern search method} \cite{hooke1961}, \textit{Brent's method} \cite{Brent1971AnAW}, and the {\em Nelder-Mead simplex method} \cite{Nelder1965}. The Nelder-Mead simplex method in particular has been frequently utilized as a local search method within hybrid metaheuristics and memetic algorithms for finding solutions to SNEs via global optimization \cite{pei2019,Ramadas2015,ouyang2009,SACCO20115424}. The books \cite{Brent1973a,Conn2009} provide a nice introduction to methods for derivative-free optimization.

% The rest of Section \ref{optimizationmethods} will focus on introducing metaheuristics which have been utilized to find solutions to SNEs via global optimization.% are discussed in Section \ref{optimizationmethods} of this paper.

% \subsection{Spectral gradient methods}

\subsection{Global search / global optimization methods}\label{metaheuristics}
Global optimization is a well developed field, and There are many ways to classify global search methods as shown in article \cite{Stork2020}, but in this paper we decide to classify them into three main categories: (1) \textit{exact methods}, (2) \textit{single-solution-based methods}, and (3) \textit{population-based methods}. We consider \textit{exact methods} to be methods which are guaranteed to find a solution via exhaustive search, we consider \textit{single-solution-based methods} to be methods which search the solution space with a single agent at a time ($\text{population}=1$), and we consider \textit{population-based methods} to be methods which search the solution space simultaneously with many agents ($\text{population}\gg1$). Since much of the recent literature discussing solution techniques for SNEs has focused on metaheuristcs, we focus our attention on the two main classes of metaheuristics: the single-solution-based and population-based methods. Hybrid methods, or methods which combine metaheuristics with local search algorithms, are not explicitly separated from the single-solution based methods and the population-based methods in this survey as many of the metaheuristics mentioned in these sections can easily be hybridized if desired. For example, see article \cite{pei2019} which hybridizes the \textit{Continuous Variable Neighborhood Search} (C-VNS) metaheuristic with the Nelder-Mead method for local search. For readers who are interested in learning more about global optimization methods and applications, we refer the interested reader to the books \cite{Horst2000,Floudas1999,horst1995,pardalos2002b,Floudas1990-mw}.

\subsubsection{Exact methods}\phantom{}\vspace*{0.1cm}

\noindent
Exact methods for global optimization guarantee that a solution will be found within a specific tolerance (if one exists) by exhaustively searching the solution space. Exact methods for global optimization are typically too computationally expensive to be applied in practice, especially in cases where the objective function is highly multimodal or the number of dimensions is very large. The \textit{Branch-and-bound algorithm} is a well known exact method for global optimization \cite{lawler1966}.

\subsubsection{Single-solution-based methods}\phantom{}\vspace*{0.1cm}

\noindent
Single-solution-based methods explore the solution space with a single agent who makes decisions about which path to take based upon 
%where to go next by following 
a predefined set of rules %which are 
typically based upon the objective function values at the points which the agent has explored so far. This section introduces  single-solution-based methods which have been utilized in literature to search for solutions to SNEs through global optimization.
% \section{Optimization techniques to find all solutions to a SNE}
%we may need short introduction here

% \noindent
% \subsection{Continuous greedy randomized adaptive search procedure}
\vspace{0.2cm}
\noindent
{\bf Continuous Greedy Randomized Adaptive Search Procedure (C-GRASP):}\/
In \cite{hirsch2007}, Hirsch {\em et al.}\/ introduced the {\em Continuous Greedy Randomized Adaptive Search Procedure} (C-GRASP) which extends Feo and Resende's {\em Greedy Randomized Adaptive Search Procedure} (GRASP) \cite{Feo1995} for discrete optimization to the domain of continuous global optimization. The C-GRASP metaheuristic utilizes multistart and a derivative-free stochastic local search method to perform global optimization. In \cite{hirsch2009}, Hirsch {\em et al.}\/ utilize an enhanced version of the original C-GRASP metaheuristic proposed in \cite{hirsch2007} to find multiple solutions to Eq.~(\ref{eq:2}) via global optimization. Each iteration of C-GRASP is initialized with an initial coarse discretization of the search space and an initial random solution vector $x$. Each iteration of C-GRASP consists of a series of construction-local improvement cycles where the output of the construction phase is used as the input to the local search phase, and where the output from the local search phase is used as the input for the next iteration's construction phase. As the iterations progress, the discretization of the search space becomes increasingly dense as C-GRASP converges to a minimum. After reaching a maximum number of multi-start iterations, the C-GRASP metaheuristic returns all the solutions to the SNE that it found.

%% CUT OUT BELOW TO REACH PAGE LIMIT
During C-GRASP's construction phases, C-GRASP utilizes greediness and randomization to create a diverse set of quality initial solutions from which to start local search. Each construction phase takes an initial feasible solution $x$ as an input, and starts by letting all coordinates of $x$ be unfixed. Then, in an iterative fashion, C-GRASP's construction phase will first conduct a line search in each unfixed coordinate direction of $x$ (while holding all other $n-1$ coordinates at their current values) to find the coordinate value along each dimension (using the current discretization) that results in the best objective function value. After conducting this line search, the best and worst objective function values found across all coordinates are saved in the variables $min$ and $max$. Then, each unfixed coordinate of $x$ is set to the best value found during the coordinate-wise line search. From here, a restricted candidate list of unfixed coordinates is formed which is comprised of the coordinates whose best objective function values found during line search were less than or equal to $\alpha \max+(1-\alpha) \min$ where $\alpha\in[0,1]$ is randomly selected at the beginning of each construction phase. Then, a dimension in the restricted candidate list is selected at random, and the corresponding coordinate value of $x$ is fixed and will no longer be moved throughout the current construction phase. By selecting a coordinate at random from the restricted candidate list (meaning the not necessarily the best coordinate is selected to be fixed) the authors preserve some randomness which they argue helps the C-GRASP metaheuristic explore more of the search space. This process repeats until all coordinates of $x$ are fixed which ends the construction phase. The final feasible point $x$ found by the construction phase will then serve as the starting point for the next derivative-free local search phase. Here, even though $x$ may be a good solution, it may not be optimal as the line search was performed along each coordinate direction. As a result, local search is then conducted starting at the feasible point $x$ found by the construction phase.
%% CUT OUT ABOVE TO REACH PAGE LIMIT

% The objective of this is to find each the feasible coordinate value along each dimension that results in the best objective function value, and then   For each coordinate, Then, Each coordinate isset starts by allowing all coordinates of $x$ to change. Specifically, each construction phase starts by allowing at an initial feasible point in the solution space, and first conducts a line search along every dimension by  

%generates an initial feasible point $x \in R^n$ which serves as the starting point for the next derivative-free local improvement phase. %C-GRASP's construction phase accomplishes thisgenerates  %Following a construction phase, C-GRASP begins a derivative-free local improvement phase from the initial point $x \in R^n$ generated by the construction phase.% is , and during C-GRASP's derivative-free local improvement phases, 

%% CUT OUT BELOW TO REACH PAGE LIMIT
During each derivative-free local search phase, C-GRASP searches in the neighborhood of the current solution for a better solution, where the neighborhood of the current solution $x$ is defined as the projection of all the current grid points onto the hyper-sphere centered at $x$ with radius equal to the current discretization level $h$. C-GRASP then
determines at which points in the neighborhood of $x$, if any, the objective function improves. Then, if an improving point is identified, it is made the new current point and local search continues from this new point. 
This local search process continues until an approximate local minimum $x'$ is reached such that taking a step in any evaluated direction would result in a worse objective function value. The $x'$ returned by each local improvement phase then serves as the starting point for the next construction phase. Specifically, the first construction phase (before any local search has been conducted) starts at a random feasible point, every construction phase after the first starts at the approximate local minimum $x'$ found by the previous local improvement phase, and each local improvement phase starts at the point generated by the previous construction phase.
%% CUT OUT ABOVE TO REACH PAGE LIMIT

The original version of C-GRASP \cite{hirsch2007} was evaluated on three SNEs: (a) a SNE arising in robot kinematics that is comprised of eight nonlinear equations in eight unknowns and which has 16 known roots \cite{Kearfott1987,Floudas1999}, (b) a SNE from the area of chemical equilibria which has ten equations and ten unknowns and which has one physical solution where all variables are positive \cite{meintjes1990}, and (c) a transformed version of the original SNE from the area of chemical equilibria which has five equations, fine unknowns, and one solution where all variables are positive \cite{meintjes1990}. The original version of C-GRASP was tested ten times on the SNE arising in robot kinematics, and successfully found all 16 solutions each time with an average runtime of 3,048s. On the original version of the SNE arising in chemical interaction and equilibrium modeling, C-GRASP successfully found the solution on eight out of ten runs with an average run time of 201.58s. On the transformed version of the SNE arising in chemical interaction and equilibrium modeling, C-GRASP successfully found the solution on all ten runs with an average runtime of 37.53s. 
The improved version of C-GRASP \cite{hirsch2009} was evaluated on four SNEs, three of which had two variables and two unknowns, and one of which had three solutions and three unknowns. Over thousands of trials, C-GRASP found all solutions to each system almost every time, and the longest average runtime for any of the problems tested was 5.06s. Other systems tested were solved in fractions of a second. 

%RePEc:eee:ejores:v:205:y:2010:i:3:p:507-521 can also be cited here.

% More specifically, 
% \begin{enumerate}
%     \item Given the SNE $F(x)$, select an initial point $x_0$ and a maximum number of iterations $k_{max}$.
%     \item For $k=0,1,2,\ldots,k_{max}$, do:
%     \begin{enumerate}
%         \item Select a forcing term tolerance $\eta^k \in [0,1)$.
%         \item If $k=0$:
%         \begin{enumerate}
%             \item Calculate the Newton-GMRES \cite{Saad1986GMRES} step $d_N$ based on the tolerance $\eta^k$. 
%             \item Proceed to step 2e.
%         \end{enumerate}
%         \item Form the local tensor model (\ref{eq:7}).
%         \item Calculate the approximate tensor step $d_T$ according to $\eta^k$ by solving one of the three methods presented for selecting $\mathcal{K_m}$.
%         \item Set $x^k+1=x^k+\lambda d$ where a linesearch strategy using the directions $d_T$ and / or $d_N$ is used to select $d$ and $\lambda$.
%         \item If $x_{k+1}$ is an acceptable approximate root of $F(x)$:
%         \begin{enumerate}
%             \item Stop.
%         \end{enumerate}
%     \end{enumerate}
% \end{enumerate}

% \subsection{Continuous variable neighborhood search}
\vspace{0.2cm}
\noindent
{\bf Continuous Variable Neighborhood Search (C-VNS):}\/
{\em Continuous Variable Neighborhood Search} (C-VNS) is a continuous extension of the \textit{variable neighborhood search} (VNS) metaheuristic proposed by \cite{MLADENOVIC19971097}, a popular method for solving discrete optimization problems. C-VNS systematically explores the solution space by first defining a set of neighborhood structures, and then repeatedly performing local search starting from a point within the corresponding neighborhoods which increase in size when improving solutions are not found. 
%after within the current  while increasing its size. 
 C-VNS was utilized to search for solutions to SNEs in \cite{pei2019}. The paper reformulates a SNE as an optimization problem and performs minimization of the resulting objective function using C-VNS. The paper defines the set of neighborhood structures to be a set of hyperspherical shells with increasing internal and external radii, so that the union of the shells covers the entire solution space. C-VNS then performs local search using the Nelder-Mead method, starting at a random point in the current neighborhood structure. The paper experimentally evaluates multiple local search methods, along with multiple ways to reformulate problem \ref{eq:1}. The paper concentrates on finding all solutions of small SNEs (2-20)%todo: check
, and the largest SNE described in the paper consisted of 1000 equations.

%  \subsection{Tabu search}
\vspace{0.2cm}
\noindent
{\bf Tabu Search (TS):}\/
Developed initially for solving combinatorial optimization problems, {\em Tabu Search} (TS) is a %non-multistart?
metaheuristic approach that can be utilized for global optimization. Starting at an initial feasible solution~$x$, TS follows a pre-defined set of rules to identify a set of ${\mathcal Y}$ solutions in the neighborhood of $x$  to explore. Let ${\mathcal N}(x)$ denote the neighborhood of $x$ such that 
${\mathcal Y} \subset {\mathcal N}(x)$. After evaluating the objective function value at each point 
$y \in {\mathcal Y}$, the point $y^* \in {\mathcal Y}$ with the best objective function value is selected as the new starting point for the next iteration of TS, even if the initial point $x$ has a better objective function value than $y^*$. By allowing moves to points with worse objective function values, TS attempts to escape the neighborhood of previously found local minima. While exploring a discrete or discretized solution space, TS stores a list of recently visited points called a ``tabu list". To ensure that TS continues to explore new regions of the solution space, TS prevents future moves to points in the tabu list. Other memory structures containing information such as the best points visited so far may also be maintained and utilized in an attempt to facilitate the performance of TS \cite{Ramadas2012}. TS continues to explore the solution space until a stopping criteria is met (in the case of SNEs, when the function value is sufficiently close to zero or upon a maximum number of iterations).

One variant of TS, named, {\em Directed Tabu Search}  (DTS), incorporates local search methods into the TS search procedure. Local search can either be conducted from the identified $y^* \in {\mathcal Y} \subset {\mathcal N}(x)$ to find a local minimizer $x^*$ from which to start the next iteration of DTS, or local search can be performed immediately from an initial starting point $x$ to find a local minimizer $x^*$ from which to select and evaluate the points $y \in {\mathcal Y} \subset {\mathcal N}(x^*)$ as described in Algorithm 2.1 of \cite{Ramadas2012}. 

Numerical results presented in \cite{Ramadas2012} showed that the described DTS method was tested on SNEs with up to $n=51$ decision variables, and that it was successful in converging on some but not all of the SNEs within $100\,n^2$ function evaluations.

% Paper \cite{Ramadas2012} has a very nice introduction, I think we could adopt pieces of it to enhance our own paper.

%, TS explores the solution space by conducting local search until a local minimum is found, and then proceeds by moving into a direction away from the current local minimum to attempt to escape the neighborhood surrounding the current local minimum before starting local search again. 

% \subsection{Simulated annealing}
\vspace{0.2cm}
\noindent
{\bf Simulated Annealing (SA):}\/
Originally proposed in \cite{Kirkpatrick1983} for  solving hard combinatorial optimization problems, {\em Simulated Annealing} (SA) is a well known metaheuristic that can be utilized to find approximate solutions to global optimization problems. SA only utilizes evaluations of the objective function during its search procedure, and can therefore be utilized in cases where information about the derivatives of the objective function is inaccessible. Although it is not always guaranteed that SA will converge to a true global minimum (some papers such as \cite{granville1994} have proved the convergence of SA under certain specific conditions), SA can be used in cases where the objective function is non-smooth or even discontinuous over the feasible region \cite{Corana1987}. 

Starting with in initial step vector $v^0 \in \mathbb{R}^n$, an initial parameter $T^0 \geqslant 0$ called the ``temperature", and an initial feasible solution $x^0 \in \mathcal{D}\subset \mathbb{R}^n$, SA attempts to generate a sequence of successive points $x^k\in \mathcal{D}\subset \mathbb{R}^n$
%$x^1,x^2,\ldots,x^k,x^{k+1},\ldots,x^K$  %such that $x^k\in \mathbb{R}^n$ and 
for $k=1,2,\ldots,K$ which approach a global minimum $x^*$ of the objective function. At each iteration of SA, 
a new candidate solution $x'$ is generated from the previous solution $x^k$ by taking random moves along each coordinate direction in turn. Each new coordinate value $x'_i$ for $i =1,2,\ldots,n$ is sampled uniformly from an interval of width $2v^{k}_i$ centered around the previous coordinate value $x^k_i$. If any new coordinate $x'_i$ makes $x'$ fall outside of the feasible region $\mathcal{D}$, then a new coordinate value of $x'_i$ is randomly generated until a feasible point is found or a maximum number of attempts is met. When a new feasible point $x'$ is found, if $\varphi(x')\leqslant \varphi(x^{k})$, we set $x^{k+1}=x'$ so that $x'$ is the starting point for the next iteration of SA. Alternatively, if $\varphi(x')> \varphi(x^{k})$, SA sets $x^{k+1}=x'$ with some probability $P(\text{acceptance})$. For example,  $P(\text{acceptance})$ can be defined as $P(\text{acceptance})=\exp(-(\varphi(x')-\varphi(x^{k}))/T^k)$ where $T^k$ is the ``temperature" parameter. Notice that if $T^k=0$, only improving points will be selected, but as $T^k$ increases, the probability that non-improving moves will be accepted increases. SA accepts non-improving moves to allow the algorithm to avoid being trapped in local minima. 
%re is a higher for large values of $T^k$ compared to $\varphi(x')-\varphi(x^{k})$, there is a high  
A new step vector $v^{k+1} \in \mathbb{R}^n$ and a new ``temperature" $T^{k+1}\geqslant 0$ can be selected at each iteration as desired. Typically, the ``temperature" $T$ and the elements of the step vector $v$ are gradually decreased as the number of iterations increases so that the algorithm can take smaller and smaller improving steps as it (hopefully) approaches a global minimum. 
% a new candidate solution $x^{k+1}$ is generated from the previous solution $x^k$ by taking random moves along each coordinate direction in turn. Each new coordinate value $x^{k+1}_i, i =1,2,\ldots,n$ is sampled uniformly from an interval of width $2v^{k+1}_i$ centered around the previous coordinate value $x^k_i$. If any new coordinate $x^{k+1}_i$ makes $x^{k+1}$ fall outside of the feasible region $\mathcal{D}$, then new coordinate values of $x^{k+1}_i$ are randomly generated until a feasible point is found or a maximum number of attempts is met. When a new feasible point $x^{k+1}$ is found, if $\varphi(x^{k+1})\le \varphi(x^{k})$

% SA performs the following search procedure:
% \begin{enumerate}
%     \item Starting at 
% \end{enumerate}

When tested against the Nelder-Mead simplex method on a variety of global optimization problems, SA was found to take 500 to 1000 times more function evaluations than the Nelder-Mead simplex method, but it more reliably found a global minimum of the objective functions it was tested on than the Nelder-Mead method \cite{Corana1987}.
Expanding upon the work presented in \cite{Corana1987}, article \cite{HENDERSON2010551} introduces a metaheuristic which utilizes SA to solve the objective function described in Eq.~(\ref{eq:modified5}) which has a penalty designed to repel SA from already found solutions. Article \cite{HENDERSON2010551} refers to this penalized objective function as a ``polarization technique".
Article \cite{EOLIVEIRA20134349} attempts to find multiple solutions to SNEs by repeatedly solving Eq. (\ref{eq:2}) with a multistart fuzzy adaptive SA algorithm. The proposed algorithm successfully found multiple solutions to SNEs of up to 20 equations and 20 unknowns ($m=n=20$) in less than four minutes when programmed in C++ and run on a computer with 2 CPU  at 2.4 GHz and 512 MB of RAM.

 \subsubsection{Population-based methods}\phantom{}\vspace*{0.1cm}
 
 \noindent
Population-based methods explore the solution space with many agents simultaneously, typically in a coordinated manner which balances exploration and exploitation. 
 %and depending on the algorithm, agents may make decisions individually or in coordination with the other agents. who may make individual decisions about which points to explore in the solution space based upon the objective function values of the points they have visited so far, or who may make decisions based upon the collective knowledge gained from the experience of the population, or who may make decisions based upon both their own experience and the experience of the group. 
 This section introduces  population-based methods which have been utilized in literature to search for solutions to SNEs via global optimization. 
\vspace{0.2cm}
\noindent
{\bf  Evolutionary Algorithms:} {\em Evolutionary algorithms}\/ are a class of metaheuristic optimization algorithms, principles of which are inspired by mechanisms of biological evolution. Often, evolutionary algorithms associate candidate solutions to biological individuals, which gradually evolve, and only the fittest survive. For example, article \cite{Geng2009} proposes an evolutionary algorithm for solving SNEs which first ranks all candidate solutions of the current population by their fitness values, and then performs mutation operations on the best candidate solutions to generate ``offspring" which will serve as the new set of candidate solutions for the next iteration of the evolutionary algorithm. Here, fitness is quantified as a mathematical \textit{fitness} function. The set of candidate solutions forms a population, that is being iteratively improved. An attractive property of some evolutionary algorithms is \textit{niching} \cite{Shir2012}, which can help find multiple solution to a SNE by 
%keep multiple optimal solutions in the population, thus, finding multiple SNE solutions. Niching techniques can be utilized to help preserve diversity by 
dividing the whole population into multiple diverse subpopulations \cite{LIAO2020113261}. We refer the interested reader to the book \cite{preuss2015multimodal} which discusses utilizing evolutionary algorithms for multimodal optimization.

A large number of research papers are devoted to the application of evolutionary algorithms for solving SNEs transformed into multiobjective optimization problems  (Eq.~(\ref{eq:multiobj})). Some of the approaches used for solving multiobjective optimization problems involve ``scalarization", that is, conversion to a problem of the form similar to 
Eq.~(\ref{summation_transformation}).
% Eq.~(\ref{eq:multiobj}). 
Evolutionary algorithms are popular as they may generate multiple solutions at the same time, which may be used to model the Pareto-front of the optimization problem. While \cite{Grosan2008} proposed to use each equation as an objective, thus forming a many-objective optimization problem, paper \cite{6849952} proposed to utilize a multiobjective evolutionary algorithm called MONES to search for solutions to a SNE reformulated as Eq.~(\ref{eq:multiobj2}) where all optimal solutions of the SNE would be Pareto optimal solutions of Eq.~(\ref{eq:multiobj2}). 

% \vspace{0.2cm}
% \noindent
% {\bf Memetic algorithms:}
% \cite{LIAO2020113261}

% \subsubsection{Genetic algorithms}\phantom{}\vspace*{0.1cm}
\vspace{0.2cm}
\noindent
{\bf  Genetic Algorithms (GAs):} %{\em Genetic algorithms.} 
{\em Genetic algorithms} (GAs) are one of the most commonly utilized types of evolutionary algorithms. {\em Memetic algorithms} (also known as \textit{genetic local search algorithms}) are a class of GAs which utilize a local search technique to attempt to improve the quality of the approximate solution(s) found. Article \cite{LIAO2020113261} discusses utilizing memetic algorithms for searching for solutions to SNEs. %, One commonly utilized class of {\em evolutionary algorithms} are {\em genetic algorithms}. % are generally considered to be a specific class of {\em evolutionary algorithm}.
Paper \cite{Silva2014} applied the {\em Random-Key Genetic Algorithm} (RKGA) for finding solutions of a box-constrained SNE, i.e. such that their solutions $x^* \in S$, where $S = \{ x = (x_1, x_2, \ldots,  x_n) \in {\mathbb R}^n: l_i \leqslant x_i \leqslant u_i  \}$. The RKGA was originally proposed for solving discrete optimization problems \cite{Bean1994}. At each iteration of the evolutionary process, RKGA computes the value of the fitness function of its chromosomes; after which, the most fit chromosomes are placed into an ``elite" group, and remaining chromosomes are considered to be ``non-elite". Then, the next population is created by combining all the ``elite" chromosomes, a small group of new random chromosomes, and another small group of new chromosomes which are generated by performing crossover between ``elite" and ``non-elite" chromosomes. There are many different variations of the RKGA, some of which which differ from the original algorithm by allowing chromosomes to have more than one offspring, and others by modifying the crossover operation so that ``parents" are taken from the entire population. The RKGA also contains an encoder and a decoder which encode and decode the solution of the optimization problem to the chromosome representation. In \cite{Silva2014}, the encoder represents a solution as a random vector with each element $\chi_i$ lying in the interval [0,1], and the decoder ``rescales" this random vector by calculating $x_i = l_i + \chi_i\,(u_i - l_i)$. In addition to the original RKGA procedure, article \cite{Silva2014} proposes to utilize a local improvement procedure to attempt to find a better solution in the neighborhood of the current solution. The gradient-free local search method utilized in \cite{Silva2014} is based on the method introduced in \cite{RePEc:eee:ejores:v:205:y:2010:i:3:p:507-521}. Article \cite{Silva2014} evaluates the method on small-scale problems from kinematics, chemical engineering, and other fields \cite{Song2020}. Article \cite{kuri2003} utilizes a GA to solve a SNE transformed into a constrained optimization problem. The constraints are handled by adding a penalty into the objective function which penalizes solutions for which the constraints are not satisfied. 

GAs have also been utilized within hybrid algorithms for solving SNEs. For example, article \cite{el-shorbagy2020} attempted to solve SNEs reformulated as a single objective optimization problem (Eq. (\ref{eq:2})) by using the novel {\em hybrid-GOA-GA algorithm} which combines the {\em grasshopper optimization algorithm} (GOA) with a GA. The {\em hybrid-GOA-GA algorithm} was tested on eight benchmark SNEs of up to ten equations and unknowns, and on the largest SNEs the {\em hybrid-GOA-GA algorithm} converged to a solution in less than 15 seconds on average. %Although the experiments were run on , and the {\em hybrid-GOA-GA algorithm} generally converged to solutions more reliably and quickly than the original GOA and GA algorithms it was compared against. The {\em hybrid-GOA-GA algorithm} converged in under 

% \subsubsection{Differential evolution}\phantom{}\vspace*{0.1cm}
\vspace{0.2cm}
\noindent
{\bf Differential Evolution (DE):}
% {\em Differential evolution.}
\noindent
Similarly to GAs, algorithms based on {\em differential evolution} (DE) work by generating a population of candidate solutions which get iteratively improved by combining candidate solutions from the previous population. New candidate solutions are generated by subsequently applying mutation and crossover operations to solutions from the previous generation; then, the solutions are compared with their parents and are selected and added to the next population if their fitness function value is better.  

Article \cite{tawhid2020hybridization} proposes to hybridize the {\em Grey Wolf Optimization} (GWO) \textit{algorithm} with a DE algorithm; the algorithm is utilized to search for solutions to a SNE reformulated as an unconstrained optimization problem. The algorithm was evaluated on small-scale SNEs and was able to find multiple solutions to SNEs without using an objective function penalty (Eq. \ref{eq:modified}) since the population of candidate solutions can potentially contain multiple solutions to a SNE found simultaneously in a single run. Article \cite{HE2019104796} proposes to explore new candidate solutions using a fuzzy neighborhood technique (based on Euclidean distance). Paper \cite{Gong2020c} also reformulates a SNE as a global optimization problem, and proposes a repulsion-based adaptive DE algorithm which utilizes additive and multiplicative repulsion techniques (cf equations \ref{eq:2} and \ref{eq:modified51}) along with adaptive control of DE parameters. Article \cite{WU2021106733} utilizes niching techniques with DE in order to search for multiple solutions to a SNE simultaneously in a single run. It has also been proposed to apply k-means clustering to the set of candidate solutions, and to treat each cluster as a separate population to preserve the diversity of the solutions \cite{Gao2020}. Article \cite{Gao2021} reformulates a SNE as a multiobjective optimization problem and attempts to solve it using a two-phase EA. The first phase attempts to maintain the population diversity by combining a multiobjective optimization technique with a niching technique, and the second phase attempts to detect promising subregions within the solution space and explore them by using DE to perform local search. Niching is also used with DE in \cite{LIAO2020105312}. Article \cite{JI2021204} applies DE with modified crossover to a SNE transformed into a tri-objective multiobjective optimization problem.

%\cite{Rovira2005}

% \subsubsection{Particle swarm optimization}\phantom{}\vspace*{0.1cm}

\vspace{0.2cm}
\noindent
{\bf Particle Swarm Optimization (PSO):}
{ \em Particle Swarm Optimization} (PSO) algorithms generate a set of candidate solutions (``particles") that are moving in the search space in accordance to some movement law. Typically, each particle has a velocity that controls its movement \cite{ParsopoulosV2010}.
%MNV MNV MNV MNV \cite{parsopoulosV2010}
%Konstantinos E. Parsopoulos and Michael N. Vrahatis. 2010. Particle swarm optimization and intelligence: Advances and applications, Information Science Publishing (IGI Global), Hershey, PA, USA.
Particles' movement directions are iteratively updated based on the best solution of the individual particle, and the current best solution of all particles. Article \cite{ouyang2009} reformulates a SNE as an optimization problem of the form Eq. (\ref{summation_transformation})
%to the sum of absolute values %todo: find equation #
and searches for solutions to it using a hybrid PSO algorithm. The hybrid PSO algorithm utilized in article \cite{ouyang2009} is a combination of PSO with the Nelder-Mead local search method. Here, some of the particles update their positions according to movement equations, and others update their positions using the Nelder-Mead method. The paper evaluates the method on small SNEs with less than seven equations, and looks for only one solution. Paper \cite{ibrahim2019} proposes a hybrid algorithm combining PSO and Cuckoo search for solving SNEs. The reformulation utilized in article \cite{ibrahim2019} is the square root of the objective function shown in Eq. (\ref{summation_transformation}) with $\alpha=1$ and $p=2$. Article \cite{li2015} utilizes PSO to minimize an objective function of the form shown in Eq. (\ref{summation_transformation}) where $\alpha=1$ and $p=1$.
%solve problem \ref{eq:2}, where $p = 1$; 
The algorithm utilized in \cite{li2015} is a modified PSO algorithm with a new expression for finding the inertia weight. The algorithm is used to find only one solution, and is tested on small SNEs. Paper \cite{turgut2014chaotic} proposes utilizing {\em Quantum behaved Particle Swarm Optimization} (QPSO) to solve a SNE reformulated as Eq.~(\ref{eq:2}). The main difference between the QPSO algorithm proposed in \cite{1330875} and the traditional PSO algorithm is that the QPSO algorithm utilizes principles from quantum mechanics while the traditional PSO algorithm utilizes principles from Newtonian mechanics.
%instead of 
%, as opposed to 
%Newtonian mechanics as used in the traditional PSO algorithm. 
% Besides QPSO, \cite{1330875} changes pseudo-random number generator to chaotic map. 
Additionally, the method presented in article \cite{1330875} changes the pseudo-random number generator to a chaotic map. 
The resulting algorithm is evaluated on small-size problems, and their experiments show that the proposed method tends to converge to solutions faster than the ordinary PSO algorithm. Article \cite{mo2009} aims at solving a SNE by minimizing an objective function of the form shown in \ref{summation_transformation} with $\alpha=1$ and $p=2$. The proposed method
%after transforming it to an optimization problem ed to square root of Eq.~(\ref{eq:2}) with $p=2$ and proposes to 
combines the standard PSO algorithm with {\em Powell's conjugate direction} (CD) method which can be used to find a local minimum of a function. The resulting method is used to find one solution and is tested on SNEs of 5-10 dimensions. PSO was also utilized to search for solutions to SNEs in %applied for SNE solutions in 
%\cite{jaberipour2011} and \cite{amaya2011real}.
\cite{amaya2011real,jaberipour2011}.
MATLAB code for PSO and a modification of PSO named ``Unified PSO" that aggregates its local and global variants~\cite{ParsopoulosV2007} can be found in~\cite{ParsopoulosV2002,ParsopoulosV2010}. PSO has also been hybridized with insights from memetic algorithms as discussed in article \cite{petalas2007memetic}.
%combining their exploration and exploitation abilities without imposing additional requirements in terms of function evaluations
%Parsopoulos K.E., Vrahatis M.N., Parameter selection and adaptation in unified particle swarm optimization, Mathematical and Computer Modelling, 46(1-2), pp.198-213, 2007.

%\cite{grailoo2011solving} %cannot find this paper

\vspace{0.2cm}
\noindent
{\bf Spiral Optimization (SPO):} Initally introduced in articles \cite{Tamura2011b, Tamura2011a},  {\em spiral optimization} (SPO) \textit{methods}\/  are multistart methods for global optimization which are inspired by spiral phenomena in nature. Article \cite{Tamura2011a} initially proposed a SPO method for solving 2-dimensional continuous optimization problems, and article \cite{Tamura2011b} expanded upon this work to introduce a SPO algorithm for $n-$dimensional continuous optimization problems. The $n-$dimensional SPO algorithm is constructed based on rotation matrices defined in the $n-$dimensional space. The iterative SPO algorithm explores the solution space with a population of solutions simultaneously, each of which explores the solution space by moving in a tightening spiral centered around the best solution found so far. Upon initialization, the spiral dynamics optimization algorithm randomly selects a set of starting points within the domain of interest $\mathcal{D}$, and evaluates the objective function value at each point. The point with the best objective function value is set as the initial center point which all other members of the population will spiral around. The motivation behind the spiral model is that initially when the spiral is large, the model will explore a diverse set of points across the solution space. As the number of iterations increases the spiral tightens, and the search intensifies around the center of the spiral to facilitate exploitation. One interesting aspect of the SPO algorithm is that it explores the solution space with a population of candidate solutions which move in a non-stochastic manner. 

Article \cite{Sidarto2015} presents a method which utilizes SPO and clustering to search for multiple solutions to a SNE. Instead of selecting the set of starting points for SPO in a purely random fashion, article \cite{Sidarto2015} utilizes the low-discrepancy Sobol sequence to attempt to select a set of starting points which are more likely to be uniformly spread across $\mathcal{D}$. By selecting starting points which are more uniformly spread across $\mathcal{D}$, one can facilitate the exploration of the model by powerfully leveraging the advantages presented by SPO's multistart framework. We refer the interested reader to Chapter 2 of the book \cite{Seydel2012} for information on low-discrepancy sequences and on how to generate random numbers with specified distributions. 

To attempt to find multiple roots to a SNE in a single run, article \cite{Sidarto2015} proposes a method which utilizes a clustering technique to identify different regions which are likely to contain a solution to a SNE, and simultaneously utilizes SPO to search for solutions within each cluster. The proposed method was tested on six SNEs of up to eight equations and eight unknowns, and the method demonstrated the ability to effectively find multiple solutions to each SNE in a single run in a matter of seconds.

% \subsubsection{Other nature-inspired metaheuristics}\phantom{}\vspace*{0.1cm}
\vspace{0.2cm}
\noindent
% {\bf Other nature-inspired metaheuristics:}
{\bf Other metaphor-based metaheuristics:}
{\em Cuckoo search}\/ models solutions of an optimization problem as eggs in a nest. Initially, there are multiple nests. The algorithm performs multiple iterations; at each iteration a randomly chosen ``egg" (a solution) in a random ``nest" is replaced by another solution (modeling a situation where a cuckoo lays her egg in another host bird's nest); then the fitness function of the ``nests" of solutions is evaluated, and the best nests are moved to the next generation, while a randomly chosen fraction of nests of worse quality is dropped (modeling the situation when a host bird discovers the cuckoo's egg and leaves the nest).  Randomization is often performed using \textit{Levy flights} random walk.
Cuckoo search was applied to SNEs in  
%\cite{ibrahim2019} % I described it in PSO
%\cite{abdollahi2016}, and \cite{zhang2019}. %todo: check and add more details
%
\cite{abdollahi2016,
ibrahim2019,
zhang2019}. 

{\em Monarch butterfly optimization} (MBO) algorithms, proposed in \cite{Wang2019}, are inspired by the behavior of North American Monarch Butterflies, and their autumn migration from Northern USA and Southern Canada to Mexico. In MBO, a population of Monarch butterflies is split into two sub-populations (modeling populations of butterflies in ``Land 1" and ``Land 2"). MBO then performs two operations, one of which is a migration operator which creates new butterflies which potentially replace butterflies from the previous population, and the other of which is a butterfly adjusting operator which changes the positions of the butterflies. These operations are iteratively applied to the two sub-populations until a stopping criteria is met. MBO is used to search for solutions to SNEs in \cite{ibrahim2019b}. 

Many other metaphor-based metaheuristics have been used to solve SNEs, including {\em invasive weed optimization} \cite{naidu2016solving,pourjafari2012}, {\em glowworm swarm optimization} \cite{zhou2012leader}, {\em hybrid artificial bee colony algorithm} \cite{jia2012hybrid}, {\em imperialist competitive algorithms} \cite{abdollahi2013}, {\em modified firefly algorithm} \cite{ariyaratne2019}, {\em harmony search} \cite{ramadas2013solving,Ramadas2014}, and a {\em soccer league competition algorithm} \cite{Moosavian2014} among others.

% \subsection{Chaotic optimization methods}
\vspace{0.2cm}
\noindent
{\bf Chaotic global optimization methods:} {\em Chaotic global optimization methods}\/ typically generate one or more chaotic sequences, and then map these chaotic sequences to a corresponding sequence of $P$ feasible solutions $x^p=(x_1^p, x_2^p,\ldots, x_n^p)\in \mathcal{D}\subset \mathbb{R}^n$ for $p=1,2,\ldots,P$ to the global optimization problem of concern. The set of generated feasible solutions $x^p$ are then used in some way, for example, they could be utilized as starting points for local search within a multi-start framework as shown in \cite{luo2008}. The reason why chaotic sequences are utilized as the basis for the set of starting points $x^p$ is because chaotic sequences have three key traits, namely, (a)~{\em pseudo-randomness}, (b) {\em ergodicity}, and (c) {\em irregularity}\/ \cite{yang2007}. 

Article \cite{koupaei2015} presented a two-stage metaheuristic to find solutions to a SNE by solving Eq. (\ref{eq:2}) where $\varphi^0(x)$ is of the form shown in Eq. (\ref{summation_transformation}). The authors of \cite{koupaei2015} claim that the first stage of the metaheuristic attempts to find a hyper-rectangle $S_2=I_{S_{2_1}} \times I_{S_{2_2}} \times \cdots \times I_{S_{2_n}}$ within the feasible region over which $\varphi^0(x)$ 
%as shown in Eq. (\ref{summation_transformation}) 
is unimodal. Starting with a larger feasible hyper-rectangle $S_1=I_{S_{1_1}} \times I_{S_{1_2}} \times \cdots \times I_{S_{1_n}}$ such that $S_2 \subseteq S_1$ (meaning that $\varphi^0(x)$  is not necessarily unimodal over $S_1$), the authors utilize the elements of a chaotic sequence in a process which iteratively shrinks $S_1$ for a set number of iterations to obtain $S_2$. It is important to note that though they hope to obtain a hyper-rectangle $S_2$ over which $\varphi^0(x)$  is unimodal, there is no guarantee that this will occur. After finding $S_2$, the second stage of the metaheuristic presented in \cite{koupaei2015} utilizes an $n$-dimensional expansion of \textit{golden section search} \cite{kiefer1953} to perform local search over $S_2$. Since $n$-dimensional golden section search is designed to be utilized on a hyper-rectangle over which the objective function is unimodal, stage one attempts to find this region. The authors of \cite{koupaei2015} tested their algorithm on a variety of SNEs of up to 10 variables and 10 unknowns, and were able to solve each problem in less than a second. However, they did not compare their algorithm to any other techniques capable of finding multiple solutions to a SNE, and there is little evidence to suggest that stage 1 of their algorithm is capable of consistently finding a hyper-rectangle $S_2$ such that $\varphi^0(x)$ is unimodal (as is required to guarantee that $n$-dimensional golden section search will converge to a global minimum of $\varphi^0(x)$  over $S_2$).

\subsection{Neural network-based optimization}
Various neural network-based optimization techniques have been utilized to search for solutions to SNEs. These approaches construct and train an  {\em artificial neural network (ANN)}\/ for each SNE; then, roots of the SNE may be found at the neurons' weights. An example of this technique is discussed in \cite{ZHANG20091136}, where it is referred to as a {\em Gradient Neural Network}\/ and used to search for the solution to system of linear equations. Similar approaches are presented in \cite{Li2008} and 
%it can be removed
\cite{Mathia2000}. 
In these approaches, solutions to the equations also correspond to the state vector of the ANN. These neural network based approaches are different from traditional modern application of ANNs, which are expected to be able to find solutions to unseen SNEs, after being trained on many SNEs such as \cite{DBLP:journals/corr/abs-1912-01412}.
In \cite{Margaris2007}, Margaris {\em et al}. presented a back-propagation trained four-layered feed-forward ANN framework for solving algebraic systems of $n$ polynomial equations with $n$ unknown variables. 
% CUT OUT BELOW TO REACH PAGE LIMIT
The four-layered network presented in \cite{Margaris2007} consists of: 
\begin{itemize}
\setlength{\itemsep}{2pt}
    \item[(a)] a single neuron input layer whose output synaptic weights are trained to become the components of one of the SNE's solutions,
    \item[(b)] a first hidden layer of $n$ summation neurons which generate each of the linear terms $\{x_1, x_2,\ldots,x_n\}$ and which have an identity activation function,
    \item[(c)] a second hidden layer of product units with custom activation functions which generate all of the nonlinear higher order terms of the algebraic system by multiplying the linear quantities received from the first hidden layer (for example, if a SNE had the nonlinear term $x_1^2$, a corresponding neuron will have an activation function that squares the input $x_1$),
    \item[(d)] an output layer of $n$ summation neurons that generate the right hand side of each nonlinear equation $f_i(x)=0$ of the SNE (serving as the estimate of the training error when compared to each equation's true right hand side value of 0),
\end{itemize}
where the only synaptic weights adjusted during training are those between the single neuron in the input layer and the neurons of the first hidden layer which represent each unknown variable of the system.  
%% CUT OUT ABOVE TO REACH PAGE LIMIT
%
%Facebook paper
% where the synaptic weights between the summation units of the first hidden layer and the product units of the second hidden layer have the fixed values necessary to contribute to the generation of the higher order nonlinear terms, and the synaptic weights connecting the two hidden layers to the output layer are set to the system coefficients and to zero for all terms not present in the system. 
% For example, consider a $2\times2$ system of the form 
% \begin{align*}
%     &\alpha_{11}x^2+\alpha_{12}y^2 +\alpha_{13}xy +\alpha_{14}x+\alpha_{15}y -\beta_1=0\\
%     &\alpha_{21}x^2+\alpha_{22}y^2 +\alpha_{23}xy +\alpha_{24}x+\alpha_{25}y - \beta_2 =0 
% \end{align*} 
 %
Expanding upon this work, Awaad {\em et al.}\/ utilized a new {\em Manhattan-Adam hybrid updating rule}\/ to accelerate model convergence and overcome the vanishing gradient problem \cite{AWAAD2019106452}. The main goal of the paper is to be able to use hardware accelerators that provide efficient implementation of this update rule. The new updating rule dynamically adjusts the step size during training, and rotates after a set number of iterations between using the \textit{Manhattan updating rule} and the {\em Adam updating rule}\/ to efficiently handle problems with sparse and vanishing gradients. 

Also expanding upon the work of Margaris {\em et al.}\/ \cite{Margaris2007}, Goulianas {\em et al.}\/ introduced the {\em General Back-propagation with Adapative Learning Rate} (GBALR) algorithm which allows the activation functions of the second hidden layer to be any function, including non-algebraic functions \cite{goulianas2016back}.

A related technique is called a {\em Zeroing Neural Network} (ZNN) \cite{JIN2017597}. ZNNs aim at zeroing error functions, the design of which depends on the problem. In the case of SNEs, similar to the research described above, the goal of ZNNs is to find a solution to a nonlinear system by applying gradient-based optimization methods. However, in contrast to previous papers which solve static systems, a ZNN introduces temporal dependency into its error function; that is, coefficients of SNEs may be dependent on an additional time parameter $t$. Also addressing temporal SNEs with ANNs, Zhang utilized a {\em Recurrent Neural Network} (RNN) to find approximate online solutions to a time-variant SNE emerging in robot kinematics \cite{ZHANG2006}.

%\section{Comparison of Solution Techniques}
% \section{Characteristics of methods capable of solving a SNE}

% \begin{table}[h!]
% \centering
% %\addtolength{\leftskip} {-0.5cm}
% \begin{tabular}{||c c c c||} 
%  \hline
%  Method & \# Roots & Derivative-Free & Classes of SNEs Capable of Solving \\ [0.5ex] 
%  \hline\hline
%  Evolutionary Algorithms & Many & Yes \& No & \\
%  \hline
%  Interval Methods &  &  & \\
%  \hline
%  Symbolic Computation Methods &  &  & \\
%  \hline
%  Homotopy / Continuation & All & Yes & Polynomial Systems \\ 
%  \hline
%  \end{tabular}
% \caption{Characteristics of methods capable of solving SNEs.}
% \label{table:1}
% \end{table}

\section{Synopsis and concluding remarks}
In this article, we have presented part two of a survey on methods for solving a system of nonlinear equations (SNE). In part two we have presented a comprehensive survey of techniques which can be utilized to search for solutions to a SNE via optimization. Since many of the SNEs that arise in real world applications are considered over a finite bounded domain $\mathcal{D}$, %we first formally introduce the problem of finding solutions to SNEs defined over a finite bounded domain $\mathcal{D}$, and 
we first introduced a technique which can be utilized to determine the number of solutions to a SNE that exist within $\mathcal{D}$. Then, we introduced various techniques which can be used to transform a SNE into an optimization problem, and we described a vast array of optimization algorithms which can then be used to search for solutions to a SNE after transformation. We discussed both local search algorithms and global search algorithms for optimization, and we place an emphasis on cutting edge metaheuristics for global optimization which have been shown to be particularly effective at finding one or many solutions to SNEs. 
%We then discuss additional methods which have been used to search for solutions to SNEs including methods from symbolic computation, homotopy / continuation methods, and interval methods. %Finally, we compared the strengths and weaknesses of the methods discussed in this paper, and we summarize which methods are best suited for searching for solutions to different classes of SNEs. 

Analyzing this literature has led us to conclude that for the hardest classes of SNEs, many of the most promising techniques that we currently have are metaheuristics such as those discussed in Section \ref{metaheuristics} of this paper. Although such techniques are not necessarily guaranteed to converge to a solution in finite time, many can be applied to ill-formed and challenging classes of SNEs as they only require evaluations of the objective function (and hence can be applied when the objective function is not everywhere differentiable). However, more research needs to be done towards improving these techniques. To address this need, we are currently exploring how machine learning can be utilized to enhance the effectiveness of metaheuristics for solving SNEs through global optimization. Additionally, as described in part one of this survey, we are also actively developing a new taxonomy of SNEs to facilitate the identification of new classes of tractable problems as well as the methods most capable of solving them. Furthermore, we are actively exploring methods capable of solving systems of nonlinear equations and inequalities.

In part one of this survey, we discussed methods for solving SNEs without transforming them into optimization problems. In part two of this survey, we discussed methods for solving SNEs by transforming them into optimization problems. In part three of this survey, we will present a robust quantitative comparative analysis of methods capable of searching for solutions to SNEs.

\bibliographystyle{unsrt}
\bibliography{LiteratureReview_no_template, symbolic_computation}
% \bibliography{symbolic_computation}
\end{document}